\documentclass[fleqn,usenatbib,useAMS]{mnras}

\usepackage{graphicx}	
\usepackage{amsmath}	
\usepackage{amssymb}	
\usepackage{multicol}        
\usepackage{bm}		
\usepackage{pdflscape}	
\usepackage{subfigure}
\usepackage{longtable}
\usepackage{caption}
\usepackage{caption}
\usepackage{booktabs}
\usepackage{threeparttable}
\usepackage{enumitem}

\usepackage[T1]{fontenc}
\usepackage{ae,aecompl}

\usepackage{newtxtext,newtxmath}

\title[]{The Properties and Evolutions of Starspots on Three Detached Eclipsing Binaries in the LAMOST-$Kepler$ survey}

\author[Wang et al.]{Jiaxin Wang,$^{1}$
	      Jianning Fu,$^{1}$\thanks{E-mail:jnfu@bnu.edu.cn}
	      Weikai Zong,$^{1}$ 
	      Yang Pan,$^{1}$
	      Hubiao Niu,$^{1,2}$
	      Bo Zhang,$^{1}$
	      \and Yong Zhang $^{3}$
\\  
$^{1}$ Department of Astronomy, Beijing Normal University No. 19, XinJieKouWai St., Beijing 100875, China\\
$^{2}$ Xinjiang Astronomical Observatory, Chinese Academy of Sciences No. 150, Science 1-Street, Urumqi 830011, Xinjiang, China\\
$^{3}$ Nanjing Institute of Astronomical Optics \& Technology, National Astronomical Observatories, Chinese Academy of Sciences, Nanjing 210042, China\\
}

\begin{document}
\label{firstpage}
\pagerange{\pageref{firstpage}--\pageref{lastpage}}
\maketitle

\begin{abstract}
	
The spotted detached eclipsing binary (DEB) offers insights into starspots on the binary.
Three spotted DEBs, KIC 8097825, KIC 6859813, and KIC 5527172, which were observed by the $\sl Kepler $ photometry and LAMOST spectroscopy, are studied in this work. The physical parameters of binaries are determined by binary modeling. The sizes, lifetimes, and single/double-dip ratio (SDR) of starspots are derived by starspot analysis. KIC 8097825 has large starspots. KIC~6859813 has a spot rotation period shorter than its orbital period but the system should be synchronized inferred from timescale estimation. The difference may be the result of the surface differential rotation. The KIC~5527172 has a long spot lifetime and an M dwarf component with an inflation radius. The primaries of these binaries and the secondary of KIC 8097825 have spots. Adding spotted DEBs of literature, we compare the starspots on binaries with those on the single stars. The spot sizes of starspots on 65\% binaries are smaller than the median of those on single stars. The lifetimes of starspots on binaries are consistent with those on single stars when the rotation periods are larger than 3 days. SDRs for half of the binaries are consistent with those of single star systems, while another half are smaller. The relative lifetime positively correlates with the RMS and SDR but negatively correlates with the rotation period. These relations are similar to those of spots on the single star systems. Binaries with luminosity ratios close to the unit tend to have more double dips.

\end{abstract}





\section{Introduction}

Starspots on binaries with rapidly rotating late-type components are prominent \citep{2003A&A...405..291H,2003A&A...405..303H,2017AJ....154..250L}, such as 
RS CVn stars \citep{2003A&A...405..763G,2020MNRAS.492.3647X} and W UMa stars \citep{2004MNRAS.348.1321B,2011A&A...529A..11S}. 
The properties of starspot on binary can be studied from Doppler/Zeeman Doppler images \citep{2003A&A...405..763G,2020ApJ...893..164X} and photometry \citep{2018MNRAS.474.5534O,2019ApJ...877...75P}. The former conveys 
the information of starspot distribution and constructs an image of stellar surface about brightness, temperature, or magnetic flux \citep{2005LRSP....2....8B,2009A&ARv..17..251S}. The lifetime, active longitude, and surface differential rotation can be deduced by comparing Doppler images at different epochs \citep{2002AN....323..349H,2020ApJ...893..164X}. This technique is limited to bright stars or rapid rotating stars and hardly can be observed continuously in a long term \citep{2007A&A...476..881K,2021MNRAS.501.1878X}. However, photometry provides complementary to that and can continuously monitor relative faint stars over a long duration, such as the $\sl Kepler$ mission \citep{2016RPPh...79c6901B}.

The lifetime, amplitude, and longitude evolution of the starspot can be determined from the photometric data.
 \cite{2017MNRAS.472.1618G} analyzed the lifetime and amplitude of starspot on single star with $\sl Kepler$ photometry. They found that both lifetime and amplitude decreased with effective temperature whereas their lifetime increased with the amplitude. Intensive investigations have also been made for starspots on binaries with photometric data, for instance,  \cite{2018MNRAS.474.5534O} used the {\sl Kepler} light curves to investigate the starspots on the binary KIC 11560447. Two active longitudes and their drift were found on the K~component of the binary. The drifts of active longitude indicated differential rotation at surface. Starspots on binary CoRoT~105895502 were analyzed with CoRoT light curves \citep{2019A&A...623A.107C}. One starspot is quasi-stationary short-lived and the other one presents a longitude prograde. These observational knowledge  provides constraints on stellar dynamo models \citep{2007AN....328.1111I,2007A&A...464.1049I,2011A&A...528A.135I,2015A&A...573A..68Y}.
 
A homogeneous sample of spotted detached eclipsing binaries (DEBs) is of benefit to investigate the relation between the starspots and their hosts. The LAMOST-{\sl Kepler} survey \citep{2015ApJS..220...19D,2016ApJS..225...28R,2018ApJS..238...30Z,2020ApJS..251...15Z,2020RAA....20..167F}, which performs follow-up spectroscopic observations for as many objects in the $\sl Kepler$ field as possible, offers an opportunity to the homogenous study. The orbital solutions of the binaries and the properties of starspots on them can be determined using the $\sl Kepler$ photometric data and LAMOST spectra. A series of study about spotted DEBs based on these data were carried out. \cite{2020ApJ...905...67P} used these data to study the binary KIC 8301013. They found two dominant starspots and their lifetimes were consistent with that on the single stars, which is inconsistent with the conclusion of \cite{2002AN....323..349H}, who found the lifetimes of starspots on the binary are longer. \cite{2021MNRAS.504.4302W}(hereafter Paper I) used these data to study the binary KIC 5359678. They found that lifetimes of two dominate starspots on this system were longer than those on single stars. In order to study this problem, more samples are needed. In this work, we use the $\sl Kepler$ photometric data and LAMOST spectra to investigate the properties and evolutions of starspots on three DEBs, KIC 8097825, KIC 6859813 and KIC 5527172. These binaries are short orbital period, eccentric, and low mass ratio systems, respectively, which offers insights into starspots on binaries with different orbital properties.
 
We describe the data obtained for the three binaries in Section 2. The binary modelling and starspot analysis are described in Section 3. The results are discussed in Section 4. A summary is given in Section 5. 

\section{Data}

\subsection{{\sl Kepler} photometry}

The three DEBs were observed by $Kepler$ in long-cadence (LC) mode with a sampling of 29.4 min between 2009 and 2013 \citep{2011AJ....141...83P,2011AJ....142..160S,2016AJ....151...68K}. The observational information is listed in Table \ref{tab:obs_info}. The contamination factors of photometry are ranging from 0 to 0.0067, which are provided by MAST \footnote{\url{http://archive.stsci.edu/kepler/data_search/search.php}}, meaning that the photometry of these targets is hardly contaminated by the light from nearby stars.

\begin{figure*}
	\centering
 
	\includegraphics[scale=0.45]{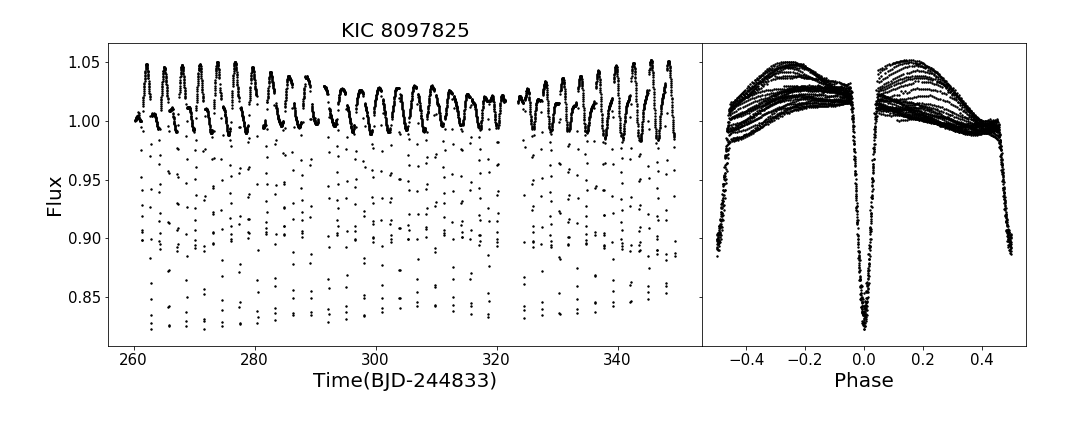}
	\includegraphics[scale=0.45]{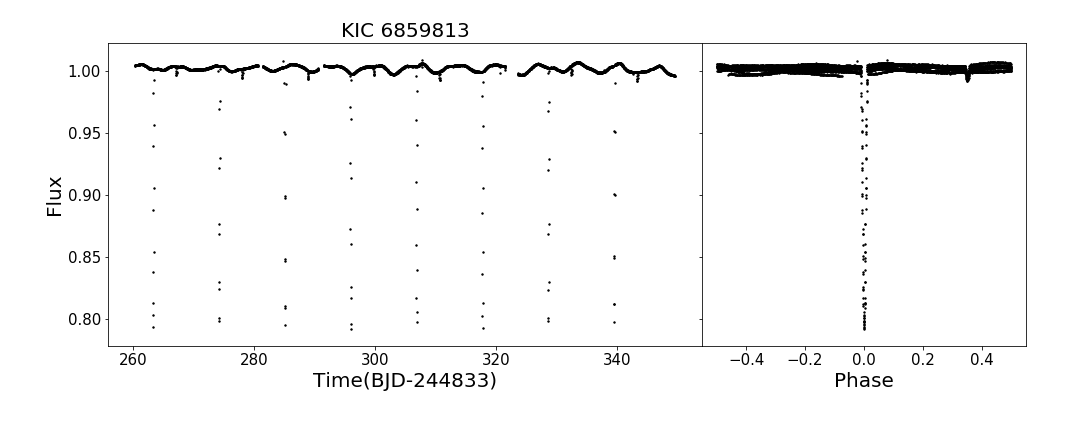}
	\includegraphics[scale=0.45]{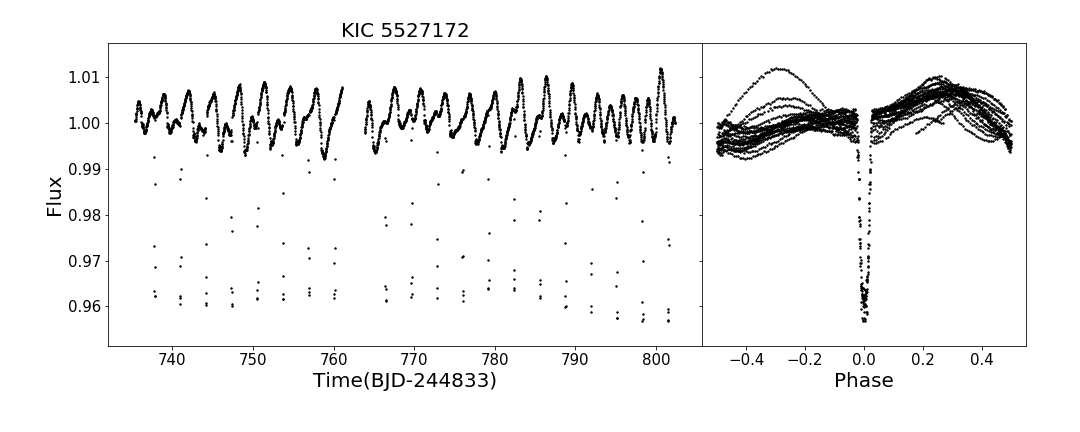}
	\caption{Observational light curves of Q3 of KIC 8097825, KIC 6859813, and KIC 5527172. The abscissa of the left panels is time and the abscissa of the right panels is phase.}
	\label{fig:LCs} 
\end{figure*}

The {\sl Kepler} photometric data are downloaded from MAST, which provides data processed by the Presearch Data Conditioning (PDC) module \citep{2012PASP..124..985S} with discontinuities, systematic trends, and outliers corrected from the raw data.
The PDC data are adopted to modelling binary light curves in this work. However, the data still suffer a long-time trend. The light curve of each quarter is then fitted by a polynomial to rectify this trend. The detrending light curves of Q3 of those DEBs are shown in the left panels of Figure \ref{fig:LCs} as an example. The time system of {\sl Kepler} data is BJD. The linear ephemerides of DEBs with orbital period and ${\rm BJD}_0$ are provided by the {\sl Kepler} Eclipsing Binary Catalog\footnote{\url{http://keplerebs.villanova.edu/}} \citep[KEBC]{2011AJ....141...83P,2016AJ....151...68K}, which are listed in Table \ref{tab:orb_par1}, \ref{tab:orb_par2}, and \ref{tab:orb_par3}. The phases of all the LC measurements are calculated with these ephemerides. The phase-folded light curves are shown in the right panels of Figure \ref{fig:LCs}.

\begin{table*}
	\caption{\centering The observation information of the three DEBs}
	\label{tab:obs_info}
	\begin{threeparttable}
	\begin{tabular}{lccccccccccc} 
		\hline
		$\sl Kepler$ ID          & RA(deg)         & DEC(deg)    &$K_{\rm p}({\rm mag})$&   Quarter  &  contamination &  & \multicolumn{2}{c}{LRS} & & \multicolumn{2}{c}{MRS} \\
		\cline{8-9} \cline{11-12}
		 (J2000.0)& (J2000.0) & & & & & & times &  SNR & & times &  SNR \\
          
		\hline
        KIC 8097825 & 291.8818 & +43.9732 & 13.283 &Q0-Q17 & 0.0039 & & 1 & 88 & & 75 & 34\\
        KIC 6859813 & 290.8067 & +42.3439 & 12.068 &Q0-Q17 & 0.0067 & &1 & 303& & 71 & 17 \\
        KIC 5527172 & 289.8181 & +40.7239 & 12.678 &Q3,Q4,Q7,Q8,Q11,Q12,Q15,Q16  &  0.0 & & 3 & 84 & & 53 & 29 \\ 
        \hline
	\end{tabular}
	
	\begin{tablenotes}
    \item[1] LRS is the LAMOST low-resolution spectrum and MRS is the LAMOST medium-resolution spectrum.
	\end{tablenotes}
    \end{threeparttable}

\end{table*}

\subsection{LAMOST spectra}

The three DEBs have been collected with LAMOST low-resolution ($R \sim 1800$) and time-domain medium-resolution ($R \sim 7500$) spectra, which had been observed by the LAMOST-{\sl Kepler} survey. The observation information is listed in Table \ref{tab:obs_info}.

The LAMOST low-resolution spectrum (LRS), which covers the wavelength range 370-900 nm \citep{1996ApOpt..35.5155W,2004ChJAA...4....1S,Cui2012,2012RAA....12..723Z}, provides the stellar atmospheric parameters by the LAMOST Stellar Parameter pipeline (LASP). The LASP uses the ULySS algorithm and the ELODIE stellar library to derive stellar atmospheric parameters \citep{Wu2010,Wu2014,Luo2004,2012RAA....12.1243L}. The LAMOST medium-resolution spectrum (MRS) constitutes the Blue arm with the wavelength range 495-535 nm and the Red arm with 630-680 nm \citep{2019RAA....19...75L,2020ApJS..251...15Z}.
The LASP is also used to measure the stellar atmospheric parameters of medium-resolution spectra \citep{2020ApJS..251...15Z}. We calculate the observation times in Barycentric Julian days and transform them into phases using the ephemerides of light curves. The spectra are chosen to derive the stellar atmospheric parameters of the primaries based on their observation phases. Five MRS of KIC~8097825 were observed in the phase from 0.5095 to 0.5270. Since the secondary eclipsing width is 0.0764 which was obtained by KEBC, these observations are in the secondary eclipsing phases. The stellar atmospheric parameters obtained from these spectra are adopted. For KIC~5527172, whose secondary eclipsing width is 0.0499, the stellar atmospheric parameters of the primary are obtained from five MRS which were observed in the secondary eclipsing phases from 0.5072 to 0.5238. The observation phases of MRS of KIC~6859813 are out-of-eclipsing. So its stellar atmospheric parameters of the primary are obtained from the LRS. These parameters are listed in Table \ref{tab:stellar_parmeter}.

The radial velocities (RVs) are extracted from MRS with the cross-correlation method \citep{2021ApJS..256...14Z}. Since the RVs have different systematic errors at different observational time \citep{2019RAA....19...75L,2020ApJS..251...15Z}, the constant radial velocity stars in the same spectrograph are used to correct the systematic variations \citep{2021ApJS..256...14Z}. We note that KIC~6859813 and KIC~8097825 are double-lined spectroscopic binaries (SB2) and KIC~5527172 is a single-lined spectroscopic binary (SB1). The RV curves are shown in Figure \ref{fig:RVs} and radial velocities are listed in appendix A.

\begin{figure}
	\centering

		\includegraphics[scale=0.35]{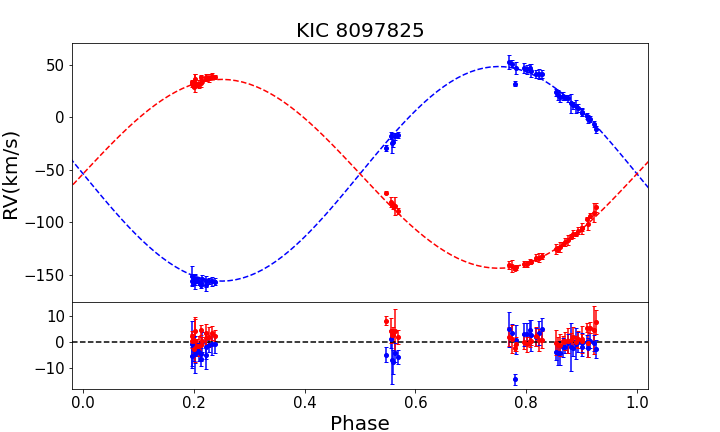}
        \includegraphics[scale=0.35]{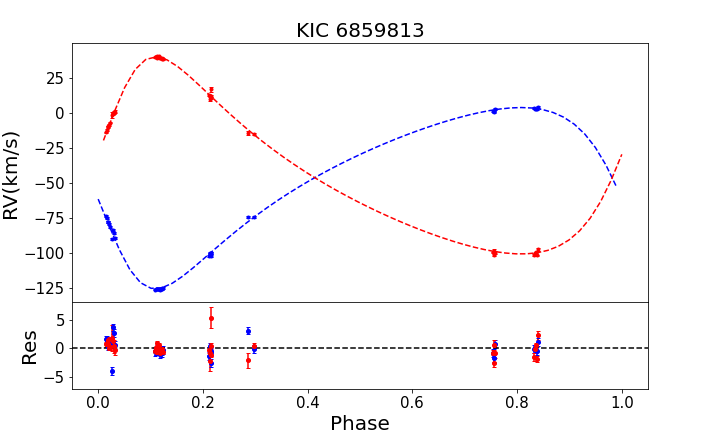}
		\includegraphics[scale=0.35]{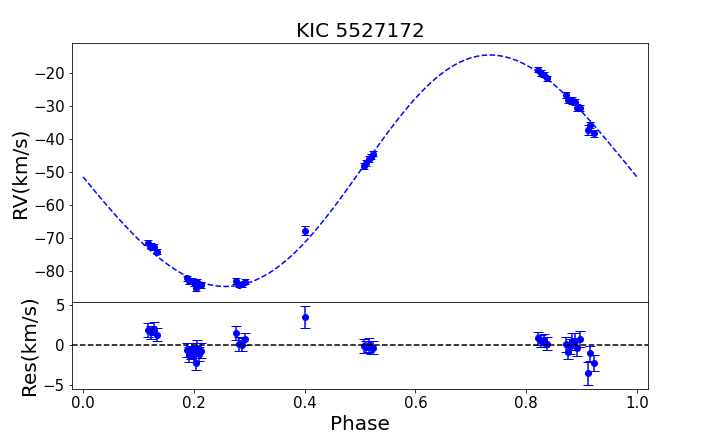}
	
	\caption{Radial velocity curves of KIC 8097825, KIC 6859813, and KIC 5527172 in the top panels. The blue circles represent the observed radial velocities of the primary and the red circles represent those of the secondary. The blue curves show the best-fitting curve of the primary and the red curves show that of the secondary. The bottom panels show the corresponding residuals.}
	\label{fig:RVs} 
\end{figure}

\section{Methods}
\subsection{Binary system parameters}

The radial velocity curves are fitted by the Keplerian orbits to derive the semi-amplitudes $K_{1,2}$, eccentricities $e$, arguments of pericenter $\omega$, and system barycentric velocities $\gamma_0$. Then, the mass ratios of SB2 are obtained to be $q=K_1/K_2$. $K_{1,2}$, $\gamma_0$, and $q$ are listed in Table \ref{tab:orb_par1}, \ref{tab:orb_par2}, and \ref{tab:orb_par3}. The best-fitting curves are shown in Figure \ref{fig:RVs}.

The light curves contain binary eclipsing signals and quasi-sinusoidal signals as shown in Figure \ref{fig:LCs}. Since the latter would affect the binary light-curve modelling, they should be modelled. The Gaussian process (GP) method \citep{2006gpml.book.....R} is adopted to model these signals. The details of how to use the GP method to simulate the quasi-sinusoidal signals have been described in Paper I. The GP kernel used in this work is as follows.
\begin{equation}
k_{\rm i,j}=A{\rm exp}\left[-\frac{(t_{\rm i}-t_{\rm j})}{2l^2}-\Gamma^2{\rm sin}^2\left(\frac{\pi(t_{\rm i}-t_{\rm j})}{P}\right)\right]+\sigma^2{\rm\delta_{ij}},
\end{equation}
where $k_{\rm i,j}$ is the covariance between the $i$th and the $j$th data while $t_{\rm i}$ and $t_{\rm j}$ are the epoches of the $i$th and the $j$th data, respectively. {Here} $A$, $l$, $\Gamma$, $P$ and $\sigma$ are free parameters. 

 The binary eclipsing signals are modelled by the PHysics Of Eclipsing BinariEs \citep[PHOEBE2.3]{2005ApJ...628..426P,2016ApJS..227...29P,2020ApJS..247...63J,2020ApJS..250...34C} code\footnote{\url{http://phoebe-project.org}}, which is based on the WD code \citep{1971ApJ...166..605W} and makes improvements which are applicable to the precise {\sl Kepler} data \citep{2013MNRAS.434..925H}. 

The light curves of each quarter are fitted individually. The light-curve fitting process is described as follows:\\
		(1) Using the GP method to modelling the out-of-eclipsing signal with a python module scikit-learn \footnote{\url{https://scikit-learn.org/stable/index.html}} \citep{JMLR:v12:pedregosa11a} and extrapolate the GP model to all the phases. \\
		(2) Subtracting the GP models from the light curves to obtain the binary signals.\\
		(3) Using the PHOEBE code to modelling the binary signals. The primary effective temperature $T_1$ is fixed to be $T_{\rm eff}$ obtained from LAMOST spectra. The free parameters are the inclination $i$, the primary mass $M_1$, the primary radius $R_1$, the secondary radius $R_2$, and the secondary effective temperature $T_2$. In addition, the mass ratios of SB2 are fixed to be the values obtained from radial velocity curves fitting and that of SB1 is a free parameter. $K_1$, $e$ and $\omega$ obtained from RV curves fitting and the primary log$g$ derived from LAMOST spectra are used as constraints for the fitting. The least-square method (LSM), whose initial values are derived by a trial of Markov Chain Monte Carlo (MCMC) sampling, is used to find the optimal orbital solution $\Theta_0$. \\ 
		(4) Performing a series of iterations. For convenience, step (3) is called as the 0 th iteration. The $i$th iteration ($i \geq 1$) is as follows:
\begin{enumerate}[label={(\alph*)}]  \setlength{\itemindent}{20pt} 
			  \item Subtracting the PHOEBE model obtained in the $(i-1)$ th iteration from the light curves. 
			  \item Using the GP method to modelling the out-of-eclipsing residuals and extrapolate the GP model to all the phases. 
			  \item Subtracting the GP model from the light curve to obtain the binary signals.
			  \item Using the PHOEBE code to modelling the binary signals to find the orbital solution $\Theta_{\rm i}$ as step (3).
			  \item Calculating the relative error between $\Theta_{\rm i-1}$ and $\Theta_{\rm i}$, $RE={\rm max}\left(\frac{|\Theta_{\rm i}-\Theta_{\rm i-1}|}{\Theta_{\rm i-1}}\right)$. If $RE\leq 0.01$, we stop the iteration. $\Theta_{\rm i}$ is adopted to be the orbital solution.

\end{enumerate}

\begin{figure}
	\centering

		\includegraphics[scale=0.35]{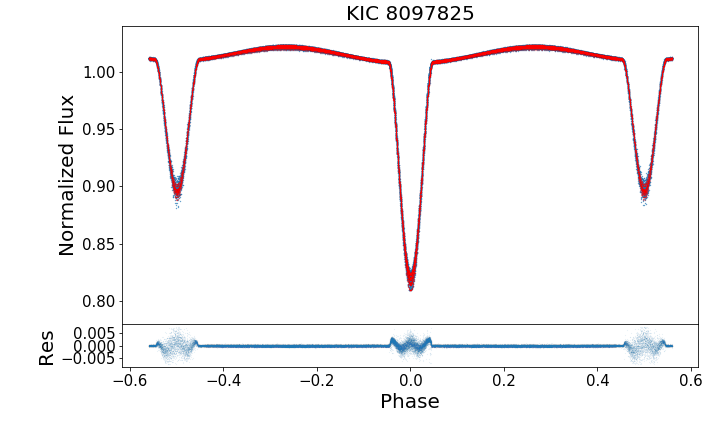}

		\includegraphics[scale=0.35]{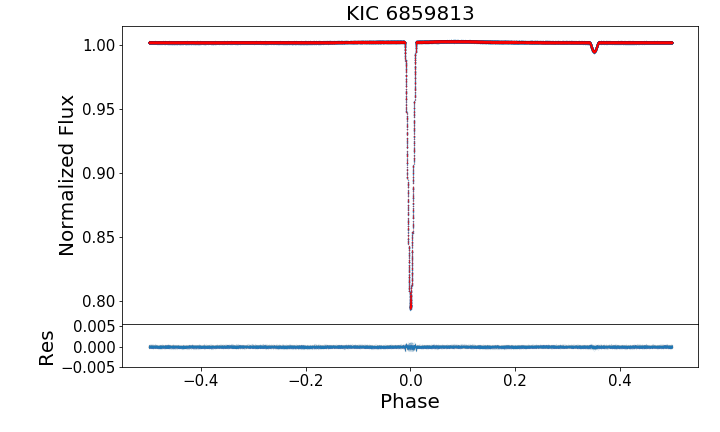}

		\includegraphics[scale=0.35]{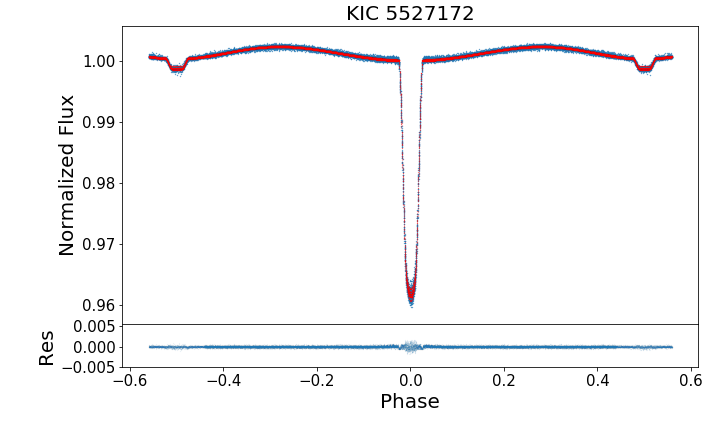}
	
	\caption{The fitting of light curves of KIC 8097825, KIC 6859813, and KIC 5527172. The upper panels show the light curves. The blue points represent the data after subtracting the quasi-sinusoidal signals. The red curves show the best-fitting models. The bottom panel shows the residuals.}
	\label{fig:LC_fit} 
\end{figure}

Finally, the orbital solutions $\theta$ and their uncertainties $\sigma$ are estimated as follows:

\begin{equation}
\theta=\frac{1}{N}\sum_{k=1}^{N}\theta_{k},
\end{equation}

\begin{equation}
s=\frac{1}{N}\sum_{k=1}^{N}s_{k},
\end{equation}

\begin{equation}
\sigma=\sqrt{\left(\frac{1}{N}\sum_{k=1}^{N}(\theta_{k}-\theta)^2\right)+s^2},
\end{equation}

where $N$ is the number of quarters, $\theta_{k}$ is the orbital solution of the $k$th quarter, $s_k$ is the uncertainty of LSM fitting of $k$th quarter. The estimation of uncertainties includes the dispersion of orbital solutions among different quarters and the mean uncertainty $s$ of LSM fitting. The orbital solutions and uncertainties are summarized in Table \ref{tab:orb_par1}, \ref{tab:orb_par2}, and \ref{tab:orb_par3}. The best-fitting models are shown in Figure \ref{fig:LC_fit}.

\begin{figure}
	\centering
	
		\includegraphics[scale=0.34]{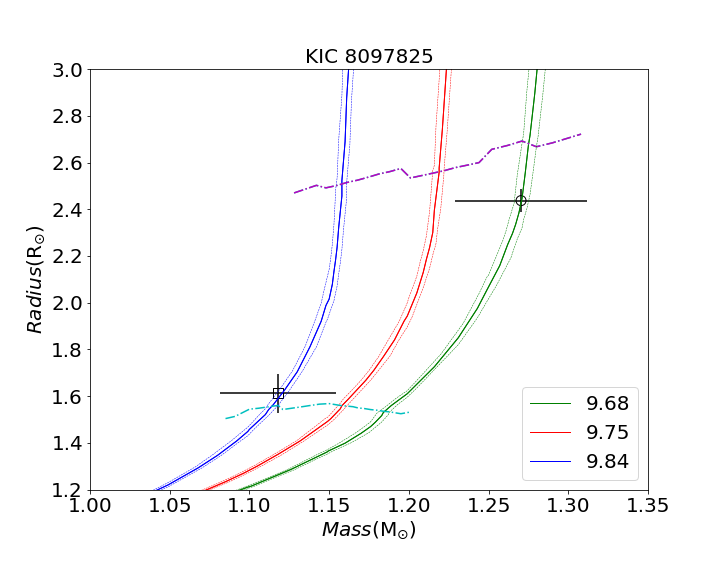}
	

		\includegraphics[scale=0.34]{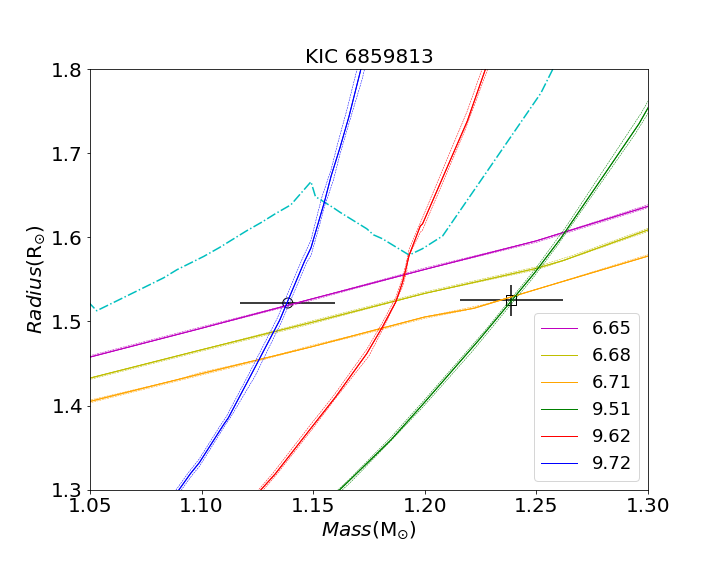}

		\includegraphics[scale=0.34]{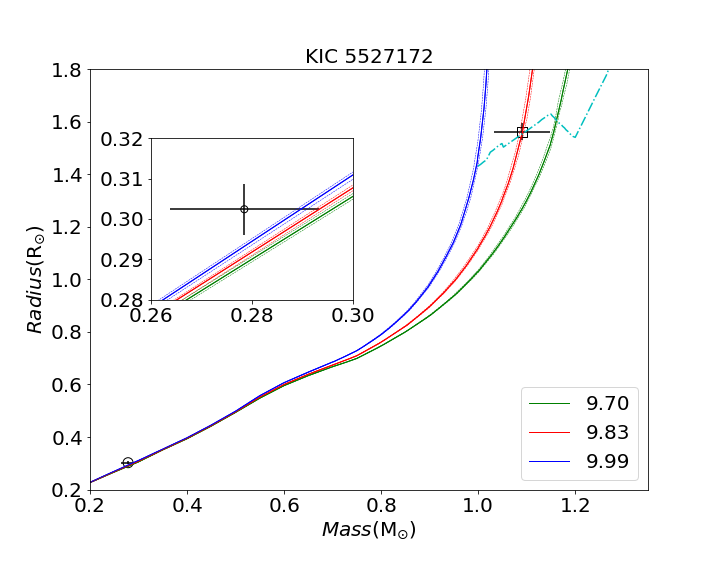}
	
	\caption{The isochrones fitting of KIC 8097825, KIC 6859813, and KIC 5527172. The square and circle are the primary and the secondary, respectively. The solid lines show the isochrones with metallicity of [Fe/H] derived from spectra. The color of line represents the ${\rm log}t$. The dashed lines show the isochrones with metallicity of ${\rm [Fe/H]}\pm \sigma_{\rm [Fe/H]}$. The cyan dot-dashed line shows the bound between the stages of main sequence and subgiant. The magenta dot-dashed line shows the bound between the stages of subgiant and red giant. To show the detail of the low-mass star of KIC 5527172, the isochrones fitting of the M dwarf is enlarged and shown on a small panel.}
	\label{fig:iso} 
\end{figure}

The ages of binaries is determined by fitting the masses and radii of the primaries and secondaries with Padova isochrones \citep{2012MNRAS.427..127B}. The fitting results are shown in Figure \ref{fig:iso}. The timescales for synchronization and circularization of binaries are calculated according to the formulas 4.42 and 4.43 of \cite{2001icbs.book.....H}, respectively.

\begin{table*}
	\caption{\centering Orbital solution and physical parameters of KIC 8097825}
	\label{tab:orb_par1}
	\begin{threeparttable}
		\setlength{\tabcolsep}{8mm}{
		\begin{tabular}{lccc} 
			\hline
			Parameter          & Primary         & System          & Secondary      \\
			
			\hline
			$T_0({\rm days})$  &                 & $2.9368523$          &                \\
			${\rm BJD}_0$      &                 & $54954.882686$       &                \\
			the width of eclipsing  &     $0.0756$  &                   &  $0.0764$                \\
			$e$                &                 & $0.0016\pm0.0011$             &                \\
			$\omega$           &                 & $84.98\pm4.97$              \\
			$A$(km/s)          & $102.36\pm0.74$ &                      & $90.09\pm0.35$\\
			$\gamma_0$(km/s)   &                 & $-53.77\pm0.16$ &                \\
			$q$                &                 & 1.136          &                \\
			\hline
			$i({\rm deg})$         &                 & $78.22\pm0.27$  &                \\
			$a({\rm R_{\odot}})$   &                 &     11.54     &                \\
			$M({\rm M_{\odot}})$   &  $1.12\pm0.04$  &                 & 1.27           \\
			$R({\rm R_{\odot}})$   &  $1.61\pm0.08$  &                 & $2.43\pm0.05$  \\
			$T_{\rm eff}({\rm K})$ &  $5820\pm30$    &                 & $5280\pm30$    \\
		    bolometric luminosity $({\rm L_{\odot} })$    &  $2.582$         &                 & $4.003$         \\
			synchronization timescale (Myr) &        &   0.85  & \\ 
			circularization timescale (Myr) &        &   320   & \\
			\hline
			gravity-darkening exponent  & 0.32        &                 & 0.32           \\
			linear-limb darkening coefficient & 0.704       &                 & 0.737          \\
			bolometric albedos         & 0.6         &                 & 0.6            \\
			\hline
		\end{tabular}}
		\begin{tablenotes}
			\item[1] $T_0$, ${\rm BJD}_0$, the widths of primary and secondary eclipsing are provided by KEBC. 
		\end{tablenotes}
	\end{threeparttable}
\end{table*}

\begin{table*}
	\caption{\centering Orbital solution and physical parameters of KIC 6859813}
	\label{tab:orb_par2}
	\begin{threeparttable}
	\setlength{\tabcolsep}{8mm}{
		\begin{tabular}{lccc} 
			\hline
			Parameter          & Primary         & System          & Secondary      \\
			
			\hline
			$T_0({\rm days})$  &                 & $10.8824223$          &                \\
			${\rm BJD}_0$      &                 & $54954.793569$       &                \\
			the width of eclipsing  &     $0.0176$  &                   &  $0.0125$                \\
			$e$                &                 & $0.3806\pm0.0006$             &                \\
			$\omega$           &                 & $126.14\pm0.08$              \\
			$A$(km/s)          & $64.60\pm0.15$ &                      & $70.27\pm0.10$\\
			$\gamma_0$(km/s)   &                 & $-46.33\pm0.08$ &                \\
			$q$                &                 & 0.919          &                \\
			\hline
			$i({\rm deg})$         &                 & $85.22\pm0.002$  &                \\
			$a({\rm R_{\odot}})$   &                 &     27.58     &                \\
			$M({\rm M_{\odot}})$   &  $1.24\pm0.02$  &                 & 1.14           \\
			$R({\rm R_{\odot}})$   &  $1.52\pm0.02$  &                 & $1.52\pm0.002$  \\
			$T_{\rm eff}({\rm K})$ &  $6130\pm10$    &                 & $6080\pm10$    \\
			bolometric luminosity $({\rm L_{\odot} })$    &  $2.856$         &                 & $2.751$         \\
			synchronization timescale (Myr) &        &   153  & \\ 
			circularization timescale (Myr) &        &   343,528   & \\
			\hline
			gravity-darkening exponent  & 0.32        &                 & 0.32           \\
			linear-limb darkening coefficient & 0.689       &                 & 0.691          \\
			bolometric albedos         & 0.6         &                 & 0.6            \\
			\hline
		\end{tabular}}
		\begin{tablenotes}
			\item[1] $T_0$, ${\rm BJD}_0$, the widths of primary and secondary eclipsing are provided by KEBC. 
		\end{tablenotes}
	\end{threeparttable}
\end{table*}

\begin{table*}
	\caption{\centering Orbital solution and physical parameters of KIC 5527172}
	\label{tab:orb_par3}
	\begin{threeparttable}
		\setlength{\tabcolsep}{8mm}{
		\begin{tabular}{lccc} 
			\hline
			Parameter          & Primary         & System          & Secondary      \\
			
			\hline
			$T_0({\rm days})$  &                 & $3.1839666$          &                \\
			${\rm BJD}_0$      &                 & $54975.414638$       &                \\
			the width of eclipsing  &     $0.0462$  &                   &  $0.0499$                \\
			$e$                &                 & $0.047\pm0.005$             &                \\
			$\omega$           &                 & $269.488\pm0.793$              \\
			$A$(km/s)          & $35.04\pm0.59$ &                      & \\
			$\gamma_0$(km/s)   &                 & $-50.05\pm0.20$ &                \\
			$q$                &                 & $0.248\pm0.017$          &                \\
			\hline
			$i({\rm deg})$         &                 & $83.99\pm0.06$  &                \\
			$a({\rm R_{\odot}})$   &                 &     10.09     &                \\
			$M({\rm M_{\odot}})$   &  $1.09\pm0.06$  &                 & $0.25\pm0.02$           \\
			$R({\rm R_{\odot}})$   &  $1.56\pm0.03$  &                 & $0.30\pm0.01$  \\
			$T_{\rm eff}({\rm K})$ &  $5880\pm30$    &                 & $3190\pm40$    \\
				bolometric luminosity $({\rm L_{\odot} })$    &  $2.652$         &                 & $0.008$         \\
		    synchronization timescale (Myr) &        &   6.5  & \\ 
			circularization timescale (Myr) &        &   884   & \\
			\hline
			gravity-darkening exponent  & 0.32        &                 & 0.32           \\
			logarithmic-limb darkening coefficient    & (0.749,0.400)     &                 & (0.873,0.596)          \\
			bolometric albedos         & 0.6         &                 & 0.6            \\

			\hline
		\end{tabular}
		\begin{tablenotes}
			\item[1] $T_0$, ${\rm BJD}_0$, the widths of primary and secondary eclipsing are provided by KEBC. 
		\end{tablenotes}}
	\end{threeparttable}
\end{table*}

\begin{table*}
	\caption{\centering The stellar atmospheric parameters of DEBs}
	\label{tab:stellar_parmeter}
	\begin{threeparttable}
	\begin{tabular}{lccccc} 
		\hline
		$\sl Kepler$ ID   & spectral type &  $T_{\rm eff}$(K) & log$g$ & [Fe/H] \\		
		\hline
		KIC 8097825       &  G3 & $5820\pm30$ & $4.018\pm0.041$  & $0.199\pm0.025$ \\
		KIC 6859813       &  F8 & $6130\pm10$ & $4.164\pm0.015$  & $0.008\pm0.008$ \\
		KIC 5527172       &  G3 & $5880\pm30$ & $4.204\pm0.036$  & $0.020\pm0.022$ \\
		\hline
	\end{tabular}
	\begin{tablenotes}
	\item[1] The spectral types are obtained from LRS. 
	\end{tablenotes}
	\end{threeparttable}
\end{table*}

\subsection{Analysis of starspot signals}

\begin{figure}
	\centering

	\includegraphics[scale=0.32]{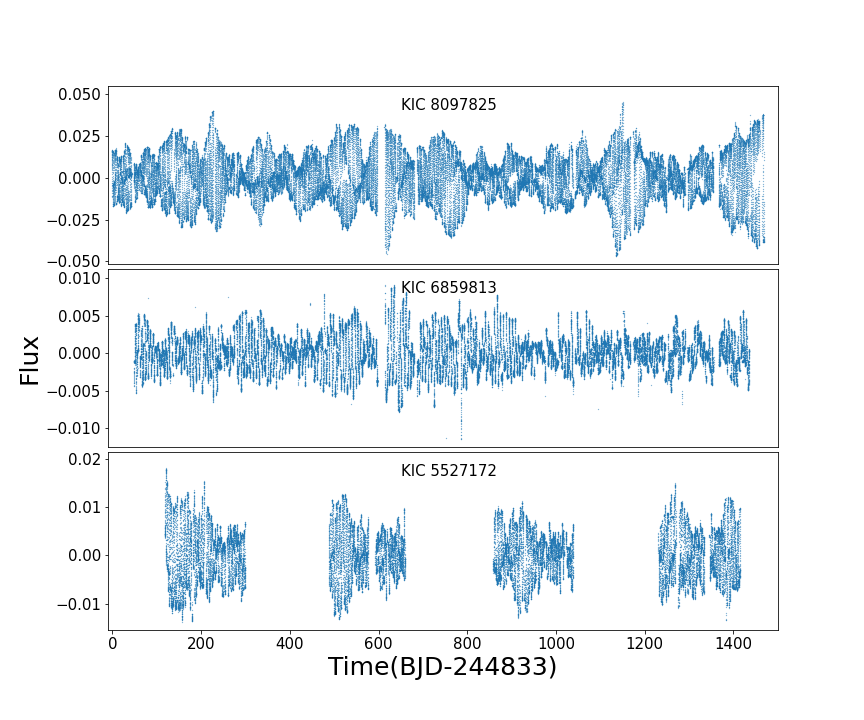}
	\caption{The starspot signals. From top to bottom panels are KIC 8097825, KIC 6859813, and KIC 5527172, respectively.}
	\label{fig:spot} 
\end{figure}

The starspot signals of the three DEBs are obtained by subtracting the binary model from the light curves. These signals are shown in Figure \ref{fig:spot} and the segments of them are shown in Figure \ref{fig:spot_segment}. It seems that the subtraction of the binary curve from the total light curve does rather little to the residual starspot light curves. Actually, the ellipsoidal variation or the reflection effect of binary would cause the variation of binary signal on the out-of-eclipsing phase, which is shown in Figure \ref{fig:LC_fit}. The residual light curve will contain these signals if the binary model is inaccurate, the residual light curve will contain these signals. Our iteration algorithm prevents the residual starspot light curves from these contaminates. 

\begin{figure}
	\centering
	
	\includegraphics[scale=0.32]{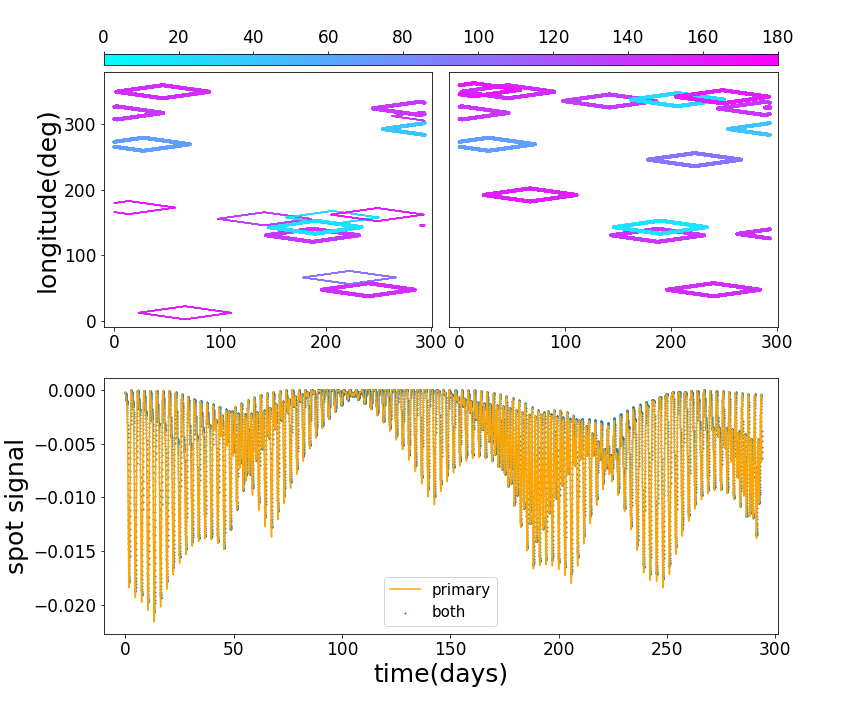}
	\caption{The simulated starspots. Two different configurations are simulated. One configuration is that both components have starspots. One configuration is that just the primary has starspots. The top panels show the distribution of starspot on these two configures, respectively. The abscissas are times and the ordinates are longitudes. The colour shows colatitude. Each rhombus represents one spot. The distance between each pair of vertical symmetrical points of the rhombus is the diameter of starspot. The rhombus shows how the size of the spot varies with time. The thick and thin lines, respectively, show spots on the primary and secondary. The bottom panel shows the simulated starspot signals. The blue and orange lines represent the spot signals of these two configurations, respectively.}
	\label{fig:mock} 
\end{figure}

The luminosity ratios of the three systems are 0.645, 0.963, and 0.003, respectively. The first two systems have similar stars that should have similar spots with similar effects. However, there is no evidence of this in their starspot signals. In fact, the light curves modulated by spots from both components of the binary could be modeled by a single star system. As a result, our starspot signals look just like many other single star systems. To illustrate it, starspots are added to both components of a binary model with parameters the same as KIC 8097825, and generate simulated light curves. Then starspots on the secondary are removed to the primary. The longitudes of these starspots are shifted by 180$^\circ$, but the other parameters do not change. The starspot signals generated by the last system could mock those of the first. The simulated starspot signals are shown in Figure \ref{fig:mock}. The starspots on both components or one component could not be distinguished by the mono-band light curve. The color index curves may be able to help with this issue. However, there is a lack of multi-band long-baseline photometric data.  

\begin{figure*}
	\centering
	
	\includegraphics[scale=0.5]{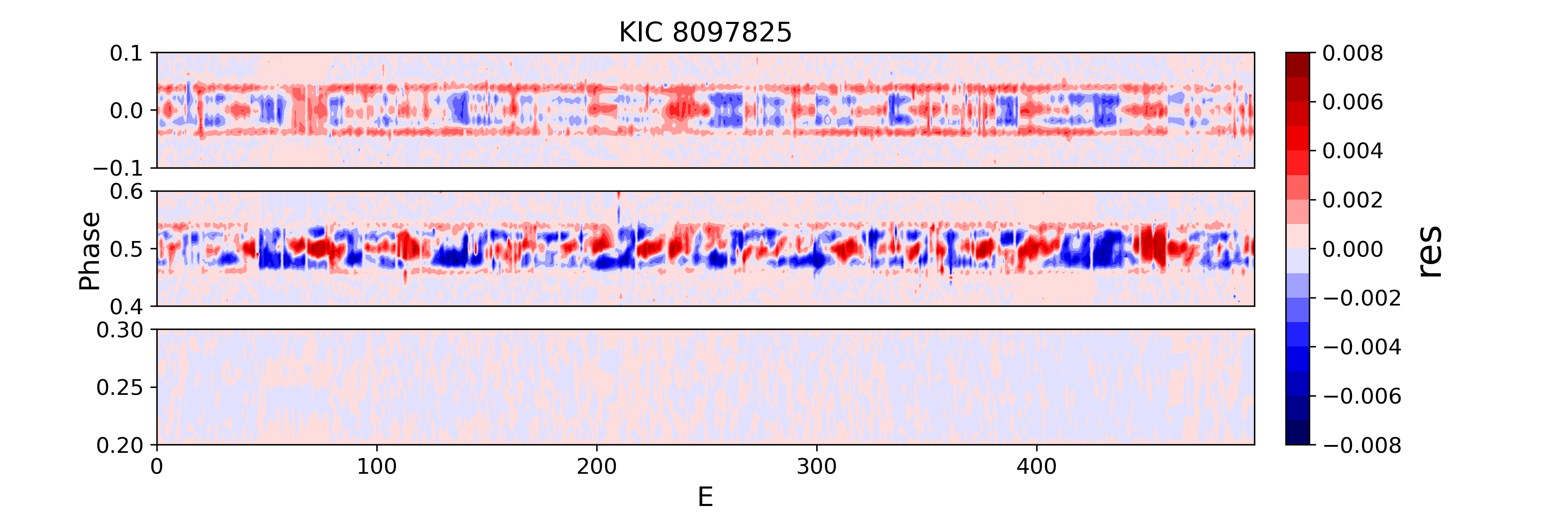}

	\includegraphics[scale=0.5]{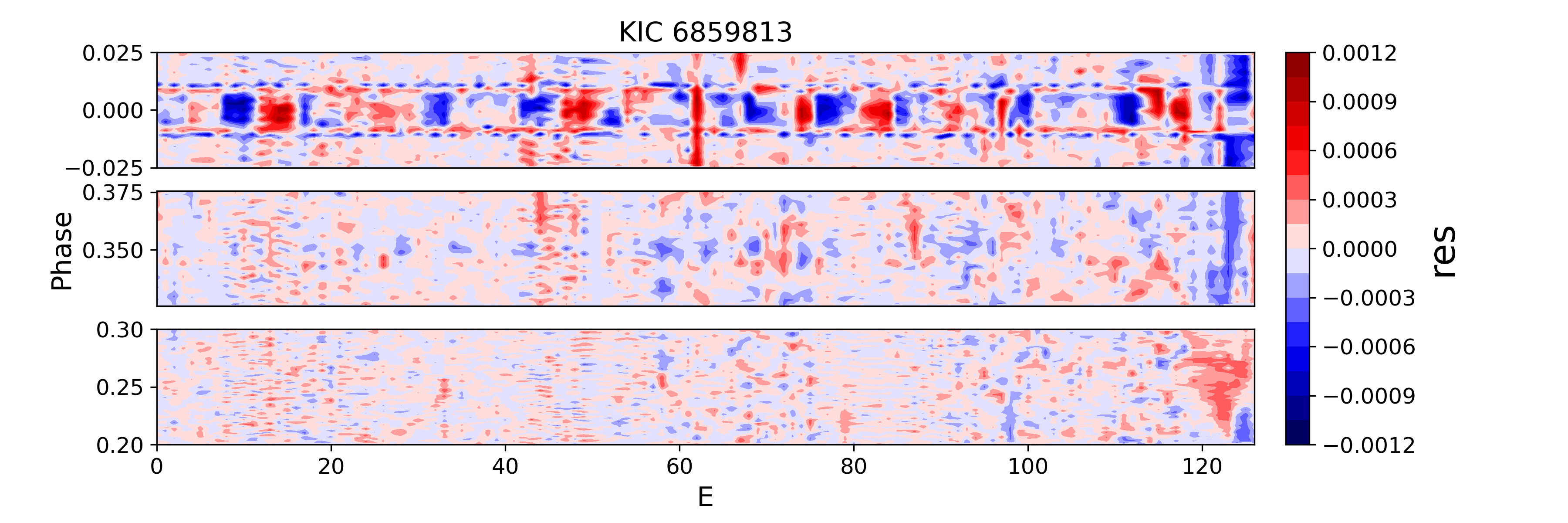}

	\includegraphics[scale=0.5]{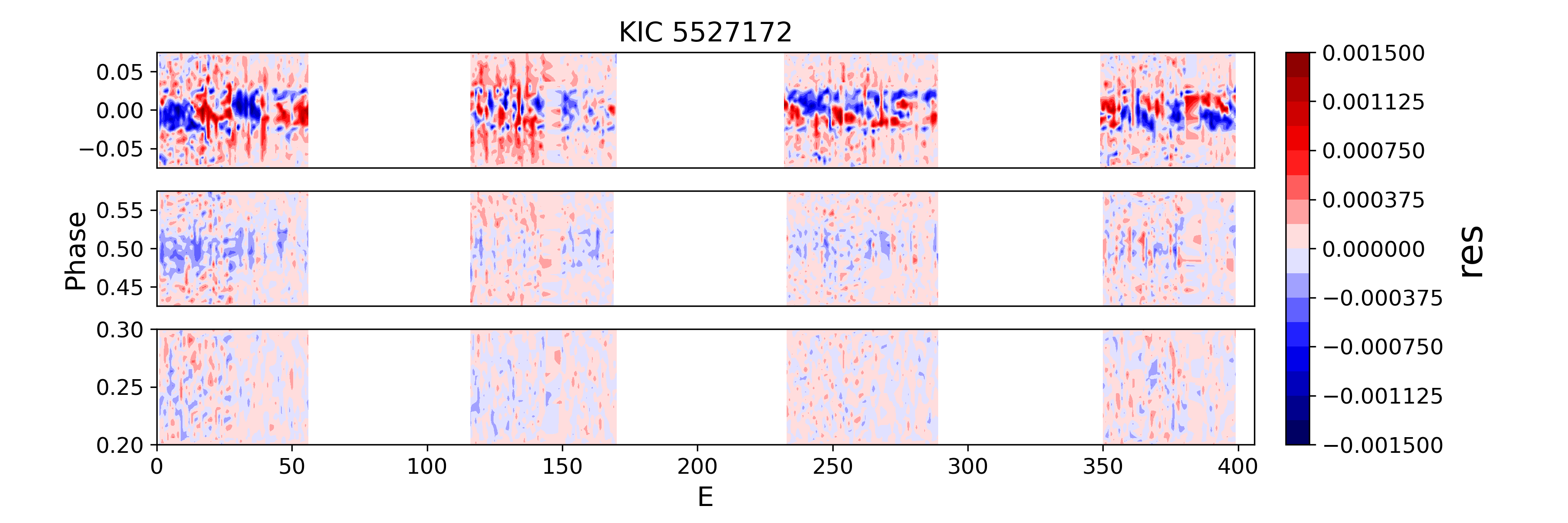}
	
	\caption{The residuals of each cycle after removing the GP and binary models from the total light curves. The abscissa is the cycle and the ordinate is the phase. The color represents the residual. The residuals near the primary eclipse, near the secondary eclipse, and in the phase of 0.2 to 0.3 are shown in the top, middle, and bottom panels for each binary.}
	\label{fig:res} 
\end{figure*}

Although the starspot light curves cannot indicate whether each star is contributing, total light curves near eclipses can provide some insight. The starspot signals would become complex during the eclipsing if starspots were eclipsed, which could not be predicted precisely by the GP model based on the out-of-eclipsing signal. The residuals near eclipse could present substructures after the GP and binary models were removed from the total light curves. Vice versa, the residuals would flatten. Therefore, these residuals could reveal which component has starspots. One extreme example is that the planet transits the starspot, the light curve becomes bright. The planet transiting events could be utilized to infer the starspot distribution on its active host star \citep{2012MNRAS.422L..72L,2020ApJ...891..103N}. The residuals of each cycle are shown in Figure \ref{fig:res}. The residuals for phases 0.2 to 0.3 are also shown in those diagrams. These out-of-eclipsing residuals can be utilized to tell if the pattern near the eclipse is considerably different from the background.

\begin{figure}
	\centering
	
	\includegraphics[scale=0.32]{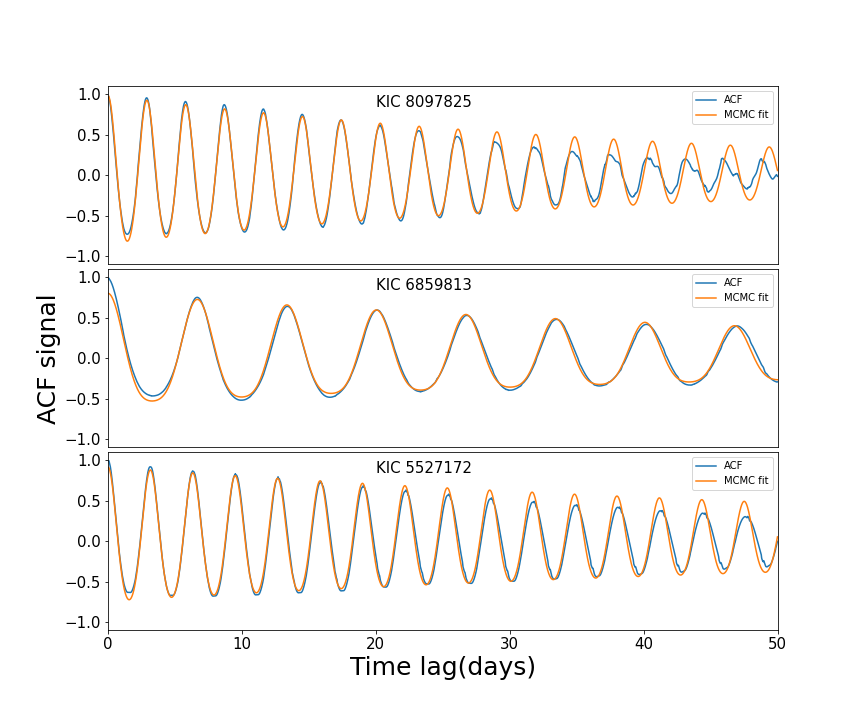}
    \vspace{-8mm}
	\caption{The ACFs of KIC 8097825, KIC 6859813, and KIC 5527172. The blue curve shows the ACF signals and the orange curve shows the best-fitting uSHO model. }
	\label{fig:acf} 
\end{figure}

To derive the rotation periods $P_{\rm rot}$ and the lifetime $\tau_{\rm AR}$ of the starspots, the auto-correlation function (ACF) method \citep{2013MNRAS.432.1203M,2014ApJS..211...24M,2015MNRAS.450.3211A} is adopted to analyze the starspots signals. The ACF describes the correlations between the light curve and itself with different time shifts (time lags). The ACF presents a peak when the time lag is at an integer multiple of the stellar rotation period. The peak decreases as the time lag increases, which is caused by the dominant starspot formation and decay. The behaviour of the ACF is similar to the underdamped simple harmonic oscillator (uSHO) at the lower time lags and becomes more complex at higher time lags which is caused by the interference of new starspot formation \citep{2017MNRAS.472.1618G}. Therefore, \cite{2017MNRAS.472.1618G} suggested using the uSHO equation as follows to fit the ACFs with time lags smaller than $2.5 \times P_{\rm rot}$. 

\begin{equation}
y(t)=e^{(-t/{\rm \tau_{AR}})}\left[A{\rm cos}\left(\frac{2\pi t}{P_{\rm rot}}\right)+B{\rm cos}\left(\frac{4\pi t}{P_{\rm rot}}\right)+y_0\right],
\end{equation}
where A and B are the amplitudes of the cosine terms, and $y_0$ the offset.
This suggestion is adopted to determine the starspot parameters in this work and the emcee code \footnote{\url{https://emcee.readthedocs.io/en/stable/}} is used to perform the fitting. However, we find that the fitting overestimates the lifetimes of starspots for stars with short rotation periods. The limit of $2.5 \times P_{\rm rot}$ is too short to obtain accurate lifetimes. Therefore, the time lag $\tau_{e}=(n+1/2)P_{\rm rot}$ is considered. $n$ is an integer and $nP_{\rm rot}$ is the time lag where the peak is lower than 1/e at the first time. It should be cautioned to use the large $\tau_{e}$, at which the ACF may be distorted from the uSHO. For the sample of \cite{2017MNRAS.472.1618G} with rotation periods about 10 days, $2.5 \times P_{\rm rot}$ is about 25 days. The time lag $\tau_{25}=(n+1/2)P_{\rm rot}$ is considered, where $n$ is an integer and $nP_{\rm rot}$ is the time lag closest to 25 days. Finally, the limit time lag is adopted to be ${\rm min}\{\tau_{e},\tau_{25}\}$. The ACFs and their fitting uSHOs of DEBs are shown in Figure \ref{fig:spot}. The rotation periods and lifetimes of starspot are listed in Table \ref{tab:spot_para}. The observation of KIC 5527172 is discontinuity and divided into four segments, as shown in Figure \ref{fig:acf}. We calculated and fitted the ACF of each segment. Then, the average fitting results are adopted.

To estimate the starspot sizes, following \cite{2017MNRAS.472.1618G}, the root-mean-square (RMS) scatter is used to represent the starspot sizes.
\begin{equation}
{\rm RMS}=\sqrt{\frac{1}{N}\sum_{i=1}^{N}y_i^2},
\end{equation}
where $N$ is the number of data points and $y_i$ is the residual of the $i$th data point. The RMS of the three DEBs are listed in Table \ref{tab:spot_para}.

\begin{figure}
	\centering

	\includegraphics[scale=0.33]{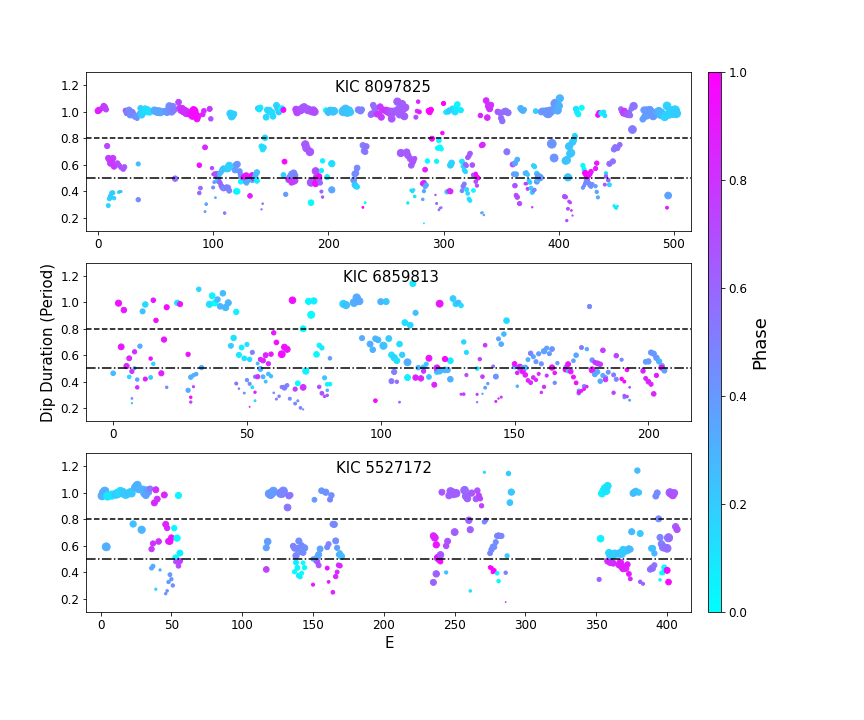}
	
	\vspace{-8mm}
	\caption{ The dip duration varies with the rotation cycle. From top to bottom panels are KIC 8097825,  KIC 6859813, and KIC 5527172, respectively. The size of the circle shows the depth of the dip. The color shows the phase of the dip. The dashed lines show the bound of double and single dips. The dot-dashed lines show the dip duration of half period.}
	\label{fig:dip} 
\end{figure}

The starspot signals present one or two dips per rotation as shown in Figure \ref{fig:spot_segment}. The simulation of \cite{2020ApJ...901...14B} reveals that the asymmetry of the complex spot distribution, rather than one or two dominant active regions, were responsible for these patterns. They also found that the number of spots, random spot distributions in position and time, spot growth and decay, and differential rotation play roles in the evolution of dips. \cite{2020ApJ...901...14B} used the duration and depth of dips and the single/double ratio (SDR) to describe the characters of single/double dips. The distance between two adjacent peaks around the dip is the duration of the dip. The flux difference between the higher peak and the dip is the depth. The dip with duration between 0.2 and 0.8 rotation is considered as double and that with duration larger than 0.8 rotation is considered as single. The logarithm of the ratio of the total duration of single dips to the total duration of double dips is SDR. The duration and depth of dips are shown in Figure \ref{fig:dip}. The SDR are listed in Table \ref{tab:spot_para}.

\begin{table*}
	\caption{ The parameters of starspots}
	\label{tab:spot_para}
	\begin{threeparttable}
	\begin{tabular}{lcccccccc} 
		\hline
		$\sl Kepler$ ID   & RMS(mag) & $P_{\rm rot}$(days) & $\tau_{\rm AR}$(days) &  $A$ &  $B$ & $y_0$ & SDR & relative location \\		
		\hline
		KIC 8097825       &  0.0133 & 2.9047 & 48  & 0.91 & 0.07 & 0.001 & 0.187 & (0,0)\\
		KIC 6859813       &  0.0024 & 6.6840 & 68  & 0.68 & 0.13 & -0.002 & -0.458 & (0,0)\\
		KIC 5527172       &  0.0051 & 3.1659 & 89  & 0.83 & 0.08 & 0.002 & -0.011 & (0,999)\\
		\hline
	\end{tabular}
	\begin{tablenotes}
	\item[1] For each binary, two relative locations are given. One is for the primary and one is for the secondary. For KIC 5527172, the secondary is much fainter, the secondary does not consider, and is labeled as 999. 
	\end{tablenotes}
   \end{threeparttable}
\end{table*}

\section{Results and Discussion}

\subsection{Results}
\subsubsection{KIC 8097825}

KIC 8097825 is a rapidly rotating binary system in our sample. The light curves of KIC 8097825 show a large amplitude of starspot modulation, obvious variations of brightness in the minima, and large widths of eclipsing phases, as shown in Figure \ref{fig:LCs}. The GP model obtained from out-of-eclipse light curves could not accurately predict the starspot signals during the eclipsing phase. Therefore, the residuals of light curves present substructures in the eclipsing phases, as shown in Figure \ref{fig:LC_fit}. The dispersion of residuals is 0.0021 and 0.0002 for the in- and out-of-eclipse phase, respectively. These large starspot signals affect the accuracy of the fitting of light curves. The uncertainties are 3.2 and 2.0-5.1 percent for the mass and radii, respectively.

We derived an age of ${\rm log}t=9.75$ for this system from the isochrone fitting in the mass-radius diagram (Figure \ref{fig:iso}). The primary and secondary lie on the isochrone tracts of ${\rm log}t=9.84$ and ${\rm log}t=9.68$, respectively. The effective temperatures determined from the isochrones are $5830\pm130~\rm{K}$ and $5120\pm90$ K, respectively. The uncertainties are propagated from the errors of masses and radii. The effective temperature of the primary is consistent with the value of $5820\pm30$ K obtained from the LAMOST MRS. The secondary is slightly cooler than $5280\pm30$ K which is obtained from the orbital solution, but they are in agreement to each other within $2 \sigma$ uncertainties. As shown in Figure \ref{fig:iso}, the secondary is evolved and close to the red giant stage, while the primary is at the end of the main sequence or somewhat leaving it. Thus the secondary is more massive, larger, but cooler than the primary. 

The timescales for synchronization and circularization are 0.85 Myr and 320 Myr, respectively. The age of this system is longer than these timescales. The orbit is circularized and synchronized. The rotation period of 2.9047 days is close to the orbital period of 2.9369 days. The relative difference is $\alpha=(P_{\rm orb}-P_{\rm rot})/P_{\rm orb}=0.011$. The rotation period is obtained from the starspot modulation. Because of the spot evolution, the period obtained from ACF is the coherence time, which is close to the rotation period but not exactly identical. This, on the other hand, is the average rotation period of starspots on various locations. This mean time could not be identical to the orbital period due to the surface differential rotation.
 
The luminosity ratio is 0.645, starspots on both components should have similar modulation on the light curves. The configuration of the system also present the similar eclipsing depth. Starspots on both components have similar probability to recognized. The residuals near the primary and secondary eclipses present obviously patterns comparing with the background as shown in Figure \ref{fig:res}, which reveals that the primary and secondary both have starspots.
These components with rapidly rotation present large starspots. The RMS of starspots of KIC 897825 is 0.0133 mag. The lifetime of starspots is 48 days. The single- and double- dips patterns alternately present. The SDR of starspot signals is 0.187. The total duration time of single dip are longer than that of double.

\subsubsection{KIC 6859813}
KIC 6859813 is an eccentric binary system. The shallow secondary eclipse of KIC 6859813 is affected by the starspot signals, as shown in Figure \ref{fig:LCs}. However, the GP method eliminates this effect, as shown in Figure \ref{fig:LC_fit}. The residuals of the light curve fitting do not show obvious substructures. The precise parameters are obtained. The uncertainties are 1.6 and 0.11-1.3 percent for the mass and radii, respectively. 

The mass-radius diagram obtains two sets of isochrones, which fit the masses and radii well, as shown in Figure \ref{fig:iso}. One set shows the binary is a young system with the age of ${\rm log}t=6.68$. The ages of the primary and the secondary are ${\rm log}t=6.71$ and ${\rm log}t=6.65$, respectively. The effective temperatures of the primary and secondary obtained from isochrones are 4750 K and 4630 K, respectively. These values are much cooler than $6130\pm10$ K from spectra and $6080\pm10$ K from orbital solution, respectively. The other set gives an old age of ${\rm log}t=9.62$. The ages of the two components are ${\rm log}t=9.51$ and ${\rm log}t=9.72$, respectively. Their effective temperatures  are $6310\pm50$K and $6150\pm40$ K, respectively. These values are slight hotter than that obtained from other methods. Since the spectra, which are used to derive the stellar atmosphere parameters of the primary, are observed out-of-eclipse, the accuracy of parameters is affected by the companion. The uncertainty of the effective temperature caused from it is about 100-150 K \citep{2018MNRAS.473.5043E}. Considering this uncertainty, the effective temperatures obtained from different methods are consistent to each other. The old-age isochrones show that the primary is on the main-sequence stage while the secondary is close to the main-sequence turnoff. 

The timescales for synchronization and circularization are 153 Myr and 343 Gyr, respectively. The age of KIC 6859813 is much longer than the timescale of synchronization. The rotation should be synchronized with the orbital motion. However, the rotation period of 6.6840 days obtained from starspot modulation signals is short than the orbital period of 10.8824 days. This may result from the surface differential rotation, which requires the relative horizontal shear  $\alpha=0.386$. This shear is near the upper limit of 0.45 of the sample of \cite{2015A&A...583A..65R}. The limit is close to the theoretical prediction from \cite{2011AN....332..933K} for model with 1.1 ${\rm M_{\odot}}$. This mass is similar with the secondary 1.14 ${\rm M_{\odot}}$ but smaller than the primary 1.24 ${\rm M_{\odot}}$. On the other hand, the starspot signal may be from the third light. There is a third body near the binary with a distance of 7.59 arcsec. Since the resolution of $\sl Kepler$ is 4 arcsec/pixel, this body may contaminate the $\sl Kepler$ photometry. The Gaia photometries are $G=11.9892$ mag and $G=18.6170$ mag for the binary and the third body, respectively. The flux ratio is 0.0022. The RMS of the spot signal is 0.0024. The contribution from the third body could not have this magnitude. There is forth body near the binary, whose distance is 10 arcsec and Gaia G band magnitude is 14.8627 mag. The contamination factor of 0.0067 given by the MAST may be the contribution from this body. However, the light curves of this body show its RMS is about 0.0005 mag. The starspot signal should be from the binary rather than other bodies. We need $vsini$ obtained from high-resolution spectra to confirm the synchronization and differential rotation.

The primary has starspots, which is revealed by the obviously patterns of residuals near the primary eclipses as shown in Figure \ref{fig:res}. Because the luminosity ratio is 0.963, starspots on both components should have comparable light curve modulation. The eclipsed region of the secondary, on the other hand, is tiny due to the high eccentricity orbit. Using residuals, it is difficult to detect if the secondary has starspots. The patterns near the secondary eclipse, however, which exhibit larger patches, are distinct from the background. The secondary appears to have starspots. Starspot signals have an RMS of 0.0024 mag. The starspots have a lifetime of 68 days. This system has more double dips as shown in Figure \ref{fig:dip}. The SDR is -0.458, indicating a longer cumulative duration of double dips.

\subsubsection{KIC 5527172}

KIC 5527172 is a single-line spectroscopic binary with an M dwarf. The secondary eclipse suffers from the starspot modulation, as shown in Figure \ref{fig:LCs}. However, the GP method obtains clean eclipsing signals, as shown in Figure \ref{fig:LC_fit}. There are no obvious substructures on the residuals of light curves fitting. The uncertainties are 6.0 and 2.1-2.2 percent for mass and radii, respectively. Since the mass ratio cannot be obtained from the RV curves, the uncertainty of mass is large. 

The age of the primary is about ${\rm log}t=9.83$, which is derived from isochrone fitting, as shown in Figure \ref{fig:iso}. The effective temperature obtained from the isochrone fitting is $5980\pm140$K, which is hotter than $5880\pm30$K obtained from LAMOST MRS. However, they are in agreement to each other within $1\sigma$. The primary is near the main-sequence turnoff or somewhat leaving it, as shown in Figure \ref{fig:iso}. The secondary does not lie on any isochrones. The radius of the secondary derived from the binary model is $6\%$ larger than that obtained from the evolution model, which predicts that the radius is about $0.285{\rm R_{\odot}}$. \cite{2018MNRAS.481.1083P} pointed out the inflation would present in both partially and fully convective M dwarfs. \cite{2018AJ....155..225K} have studied a sample of fully convective rapidly rotating M dwarfs. They found that the radii of stars with a mass range of $0.18 {\rm M_{\odot}} < M < 0.4 {\rm M_{\odot}}$ are larger than the predictions of evolution models by 6\%, on average. The percent of radius inflation obtained here is consistent with the result of \cite{2018AJ....155..225K}. This is an interesting low-mass star for studying magnetic inflation. 

The timescales for synchronization and circularization are 6.5 Myr and 884 Myr, respectively. The age of this system is longer than these timescales. Orbit is circularized and synchronized. The rotation period of 3.1659 days is close to the orbital period of 3.1840 days. The relative difference is 0.006.

The primary has starspots, as evidenced by the clear patterns of residuals surrounding the primary eclipses, as shown in Figure \ref{fig:res}. The modulation from the starspots on the secondary is weak because it is much fainter, with a luminosity of 0.003 of the primary. There is not that much of difference between the residuals near the secondary eclipse and the background. However, some clues have been discovered. The residuals at the first segment are different from the background. The negative residuals concentrate near the secondary eclipse in the next two segments, while positive residuals surround them, which are different from the background. The last segment is consistent with the background. The secondary seems like has starspots. The RMS of starspot signals is 0.0051 mag. This means that the RMS of the starspot signal caused by the secondary is about $1.5 \times 10 ^{-5}$ mag. The RMS of residuals in the out-of-eclipsing phase is about $1.7 \times 10^{-4}$ mag. Therefore, the  patterns of residuals near the secondary eclipse are caused by the fitting method rather than the effect of the spots of the secondary. The lifetime of the starspots is 89 days. The SDR is -0.011, meaning that the total duration time of the double dips is close to those of single.

\subsection{Comparing the RMS and lifetime with the single star system}

\begin{figure}
	\centering
 
		\includegraphics[scale=0.39]{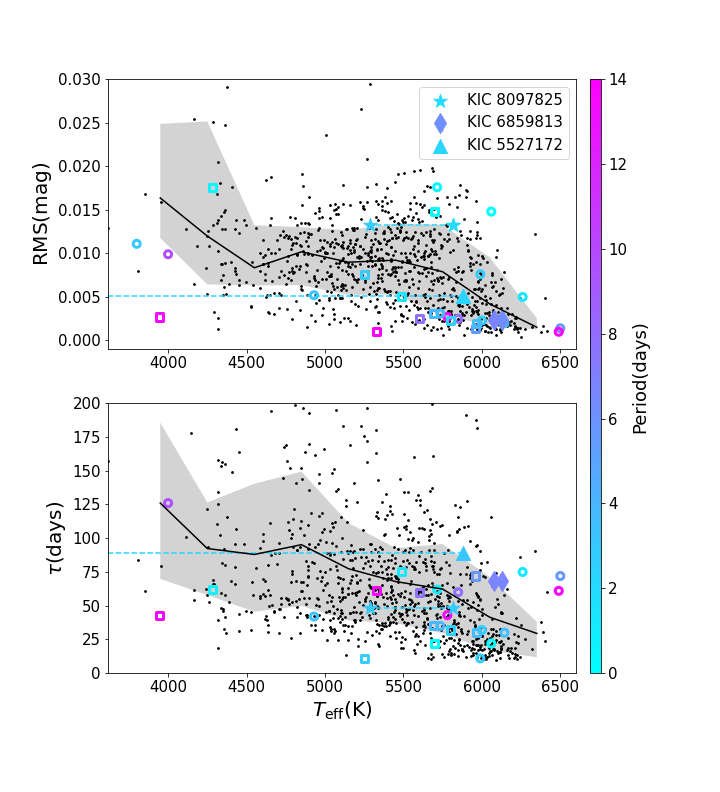}
	\vspace{-8mm}	
	\caption{Comparing the RMS and the lifetime of starspots on the binaries and single stars. The black dots are for single stars. The open red circles and squares are primaries and secondaries of binaries from literature, respectively. The filled markers are binaries in this work, respectively. The primaries and secondaries of binaries in this work are connected with dashed lines. The color represents the rotation period. The black solid lines show the medians of RMSs and lifetimes of starspots on single stars. The grey shadow shows the $1 \sigma$ region around the median.}
	\label{fig:compare} 
\end{figure}

We use the parameters of starspots of single stars in the sample from \cite{2017MNRAS.472.1618G} to compare the properties of starspots of binaries with those of single stars. Two samples were investigated by \cite{2017MNRAS.472.1618G}. Their rotation period ranges are 9.5-10.5 days and 19.5-20.5, respectively. For the majority of the binaries on this work or collected from literature, the rotation periods are shorter than 10.5 days. So the sample with a rotation period range of 9.5-10.5 days is adopted to make comparison, which are shown in Figure \ref{fig:compare}. 

The RMS of starspots of KIC 897825 is larger than most single stars with the same effective temperature. This is caused by the fast rotations. The lifetime of starspots is consistent with those of single stars. The RMS of starspot signals of KIC 6859813 is consistent with those of starspots on the single stars. The lifetime of the starspots is longer than those on most of single stars with the same effective temperatures. The starspots might locate at high latitudes, where the starspots have long lifetimes
\citep{2002AN....323..349H,2005LRSP....2....8B,2009A&ARv..17..251S}. On the other hand, the longer lifetime may be the property of starspots on the binary \citep{2002AN....323..349H}. The RMS of starspot signals of KIC 5527172 agrees with those on the single stars with the effective temperature of the primary. However, the lifetime of starspots is longer than those on most single stars with this effective temperature. The sample of single stars does not contain stars with effective temperatures near that of the secondary. We try to estimate the lifetime of starspots on those single stars by extrapolating the empirical formula of \cite{2017MNRAS.472.1618G} to that temperature. The expectation of the lifetime of starspots on the secondary is $81\pm33$ days. This is consistent with the result of our analysis for KIC 5527172. However, the extreme faint secondary has weak starspot signals comparing with that form the primary. The signals are mainly from the starspots on the primary. It seems like the starspots on this system have long lifetime.

To learn more about the differences, starspot modulated DEBs with orbital solutions obtained from $\sl Kepler$ photometry and RV curves are collected from the literature. Since the orbital solutions of those DEBs were obtained in the literature, we use the same method as Section 3.2 to extract and analyse the starspot signals of DEBs. The information and parameters of those DEBs are listed in Table \ref{tab:sample}. The RMSs and lifetimes of starspots on binaries are compared with those on single stars in Figure \ref{fig:compare}. The RMSs of starspots on 67\% binaries of our sample are lower than the medians of those on the single stars. Most of them are around or below the lower boundary of $1 \sigma$ region of single stars, as shown in Figure \ref{fig:compare}. It seems like the components of binary have smaller starspots. The lifetimes of starspots on 66\% binaries of our sample are in agreement with that on the single stars within $1\sigma$. A few primaries, whose effective temperatures are larger than 6000 K, are located above the upper boundary. However, their cooler companions lie in the $1\sigma$ region, except for one binary. Based on this small DEB sample, the lifetimes of starspots on binaries are not obviously longer than those on the single stars.

\begin{figure}
	\centering
	
	\includegraphics[scale=0.5]{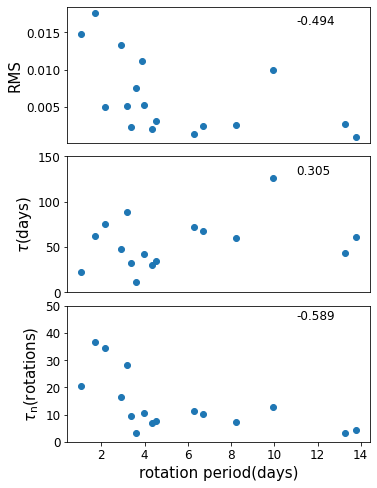}
	
	\caption{Top panel, the relation between the rms and rotation period.
	Middle panel, the relation between the lifetime and rotation period.
	Bottom panel, the relation between the relative lifetime and rotation period.
	The Pearson correlation coefficient is labelled in each panel.
}
	\label{fig:rms_tau_period} 
\end{figure}

As shown in the top panel of Figure \ref{fig:compare}, the rotation period may affect the RMS. The fast rotators have larger RMS, while the slower rotators have smaller RMS. The RMS-period diagram (top panel of Figure \ref{fig:rms_tau_period} ) shows that the RMS negatively correlates with the rotation period with a Pearson correlation coefficient of -0.494. This relation is consistent with the result of \cite{2018ApJ...863..190B}, which study the starspot signal on the single star systems based on the $\sl Kepler$ photometry. Excluding the binaries with rotation periods larger than 10.5 days, the RMSs of starspots on 65\% binaries of our sample are lower than the medians of those on single stars.

The lifetime does not present correlation with the period, as shown in the bottom panel of Figure \ref{fig:compare} and the middle panel of Figure \ref{fig:rms_tau_period}. The relative lifetime $\tau_{n}=\tau/P_{\rm rot}$ decrease with the period, which is consistent with the result of \cite{2021arXiv211013284B}, which investigated starspots on single star systems. The RMS does not correlate with the lifetime with the Pearson correlation coefficient of 0.007, which is revealed by the top and middle panels of Figure \ref{fig:rms_tau_period}. This is different from the result of \cite{2017MNRAS.472.1618G}, which shows a positive relationship with increasing lifetime and larger spots. Their result is based on the sample with a narrow period range, while our sample spans on a wide period range. Using the relative lifetime to instead the lifetime, the consistent result is obtained.

\begin{figure}
	\centering

	\includegraphics[scale=0.6]{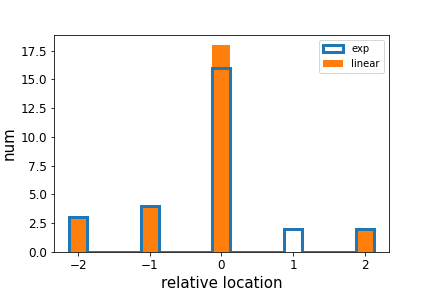}
	
	\caption{The histograms of the relative locations of the binary data points on the period-lifetime diagrams. The blue rectangles show the data obtained from the uSHO equation. The orange bars show the data obtained from the linear decay equation.}
	\label{fig:location} 
\end{figure}

In our sample, 82 percent of binaries have rotation periods shorter than 9.5 days. In order to compare the lifetime in the  corresponding range of rotation period, the sample of \cite{2021arXiv211013284B} is adopted. Single stars with rotation periods ranging from 3 to 50 days and effective temperatures ranging from less than 4000 to 6800 K are included in their rotation-lifetime diagrams. Since their data are not published, the data points of binaries are put on their rotation-lifetime diagrams. The relative locations are coded as following. -2,1,0,1,2 represent that the lifetimes are below the bottom edge, near the bottom edge, in the range, near the top edge, and above the top edge, respectively. The spot lifetimes of KIC 8097825 and KIC 6859813 are consistent with those of single star systems, while KIC 5527172 has a long spot lifetime.  The histogram of locations of all binaries is shown in Figure \ref{fig:location}. Lifetimes of starspots on binaries are consistent with those on single star systems. \cite{2021MNRAS.508..267S} found that using the uSHO equation to fit the ACF would underestimate the lifetime. They adopted the linear decay equation to fit the ACF. We also adopted this equation to fit the ACF. These fitting results are compared with the single star sample. The comparison result is shown in Figure \ref{fig:location}. These lifetimes are consistent with those on the single star systems. \cite{2021MNRAS.508..267S}  pointed out that the ACF decay can be used to obtain a lower limit of the active-region lifetimes. Therefore, this consistent may result from the underestimate of lifetime. The single star system sample, one the other hand, just includes stars with rotation periods greater than 3 days, hence the binaries with periods smaller than 3 days cannot be compared. Perhaps the effect on the starspot evolution from the companions would be detected for a closer orbit.

Comparing with single stars, the components of binary suffer tidal forces from their companions. \cite{2000AN....321..175H,2002AN....323..399H} showed that the tidal force affects the emergence of starspots and causes the active longitudes. The simulation of \cite{2007A&A...464.1049I} found that the lifetimes of starspots on active cool stars strongly depend on the details of the interactions between surface magnetic fields and large-scale convective flows, and the natures of turbulent diffusion due to convective flow patterns. Therefore, these observational results offer constraints to study the tidal effect on the convective flow patterns and surface magnetic fields.

\begin{table*}
	\caption{\centering The DEBs collected from literatures}
	\label{tab:sample}
	\begin{threeparttable}
	\setlength{\tabcolsep}{4.5mm}{
	\begin{tabular}{lccccccccc} 
		\hline
		$\sl Kepler$ ID   & $T_1$(K) & $T_2$(K) & $e$ & $q$ & $P_{\rm rot}$(days) & $\tau$ (days) &RMS(mag) & SDR &  ref \\		
		\hline

 KIC 5359678  & 6500 & 5960 & 0.0032 & 0.736842 &  6.2821 &  72 &  0.0014 & -0.6029 & 1\\ 
 KIC 8301013  & 6144 & 5966 & 0.0014 & 0.765517 &  4.3200 &  30 &  0.0020 & -0.6405 & 2\\
 KIC 3120320  & 5780 & 3945 & 0.0340 & 0.411061 & 13.2580 &  43 &  0.0027 & 0.8536 &3\\
 KIC 8552540  & 6060 & 5700 & 0.0000 & 0.678014 &   1.0653 &  22 &  0.0148 & 0.6619 & 4\\
 KIC 9641031  & 6260 & 5490 & 0.0000 & 0.764469 &  2.1649 &  75 &  0.0050 & 0.1272 & 4\\
 KIC 10987439 & 6490 & 5330 & 0.0509 & 0.655415 & 13.7840 &  61 &  0.0010 & -0.5579 & 4\\
 KIC 11922782 & 5990 & 5250 & 0.0000 & 0.556962 &  3.5884 &  11 &  0.0076 & -0.0599 & 4\\
 KIC 6525196  & 6000 & 5800 & 0.0000 & 0.870072 &  3.3817 &  32 &  0.0023 & -0.4859 & 4\\
 KIC 6781535  & 5849 & 5600 & 0.2510 & 0.862385 &  8.2340 &  60 &  0.0025 &  -0.6888 & 5 \\
 KIC 4285087  & 5735 & 5689 & 0.0000 & 0.938238 &  4.5306 &  35 &  0.0031 & -0.8186 & 5\\
 KIC 3241619  & 5715 & 4285 & 0.0000 & 0.920430 &  1.6822 &  62 &  0.0176 & 0.5821 &  6\\
 KIC 7605600  & 3800 & 3000 & 0.0013 & 0.339321 &  3.8911 & 755 &  0.0111 & 0.2793 & 7\\
 KIC 9821078  & 4000 & 3300 & 0.0314 & 0.800604 &  9.9353 & 126 &  0.0099 & -0.0518 & 7\\
 KIC 10935310 & 4930 & 3097 & 0.0439 & 0.543478 &  3.9741 &  42 &  0.0052 & -0.2898 & 8\\
		\hline
	\end{tabular}}
    \begin{tablenotes}
    	\item[] 1 \cite{2021MNRAS.504.4302W}.  2 \cite{2020ApJ...905...67P}. 3 \cite{2016MNRAS.461.2896H}. 4 \cite{2017MNRAS.468.1726H}. 5 \cite{2019AJ....158..106C}.  6 \cite{2017AJ....154..216M}.  7 \cite{2019AJ....158..111H}. 8 \cite{2017AJ....154..100H}.
    \end{tablenotes}
\end{threeparttable}
\end{table*}

\begin{figure}
	\centering
	
	\includegraphics[scale=0.42]{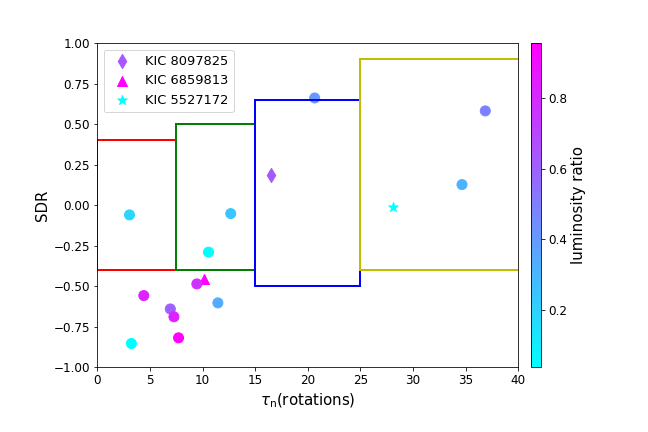}
	\vspace{-3mm}
	\caption{The relation between the SDR and relative lifetime. The color of the circle represents the luminosity ratio. The circles are binaries from literature. The other markers are binaries in this work. The rectangles show the SDR range of the single star systems of the simulation of Basri \& Shah (2020).}
	\label{fig:SDR} 
\end{figure}

\subsection{The SDR of binary system}

The SDR describes the total character of the single and double dips. The relation between the SDR and the relative lifetime is shown in Figure \ref{fig:SDR}. The SDR shows a positive correlation with the $\tau_{n}$. The Pearson correlation coefficient is 0.748. The Figure 7 of \cite{2020ApJ...901...14B} also showed the SDR peak value increases with $\tau_{n}$, when $\tau_{n}$ is small. The complexity of the spot distribution can change significantly in a few rotations for a short relative lifetime, which offers more chances to form double-dip configuration. The SDR also has a negative correlation with the rotation period. The Pearson correlation coefficient is -0.617. This is consistent with the study of \cite{2018ApJ...863..190B}.

\cite{2020ApJ...901...14B} study the SDRs of single star systems with inclinations of 30, 45, 60, and 90 degrees. Since the inclinations of our systems are larger than 78 degrees. The sample with 90 degrees is used to compare. For this inclination, they calculated the SDR for starspots with relative lifetime of 5, 10, 20, and 50. They considered cases with no
differential rotation and solar differential rotation. The solar differential rotation case is used in this study. The ranges of SDR of single star systems are illustrated in Figure \ref{fig:SDR} with rectangles. SDRs for half of binaries are consistent with those of single star systems. The SDRs of KIC 8097825 and KIC 5527172 are located in these ranges. Their starspot signals are like those on the single star systems. SDRs for another half of binaries are smaller than those of single star systems. \cite{2020ApJ...901...14B} considered that the false double dips caused by photometric errors would decrease the SDR. To avoid this effect, the dip with a depth larger than 0.0001 is adopted to calculate the SDR. The components of most these binaries have similar luminosity. The components both contribute to starspot signals if they both have starspots. The number of starspot increases compared to the single star system. It seems like a greater number of starspot would tend to have more double dips. The simulation of \cite{2020ApJ...901...14B} showed that the distributions of SDR are similar when the number of starspot is more than six. As displayed in Section 3.2, combining the starspot distributions from two stars and applying them to a single star, the signal of this star could mock the signal combined by two stars. It seems like the starspot signals from binary are similar to those from single star systems. The simulation requires large spot fill factor, synchronous rotation, sophisticated temperature factors of spot, and no differential rotation. These strict requirements are easy to break down for practical situations. Starspots that come from two stars would have a more complex configuration. The starspot signals tend to have a long total duration time of double dips. However, the relative lifetime and luminosity ratio competition make SDR cannot perfectly tell the starspot signals from binary or single star system. KIC 8097825 is one example, who both has a large $\tau_{n}$ and luminosity ratio. We need long-baseline high-quality multi-band photometric data to learn the detail of starspot, which offers temperature information that is as important as the luminosity information.

\section{Summary}
Three detached binaries with starspot modulation are studied in this work. Combining the $\sl Kepler$ photometry and the LAMOST spectra, the orbital solutions of the three DEBs are obtained. Using the isochrones to fit the masses and the radii of the primaries and secondaries, the ages of binaries are determined. The effective temperatures derived from isochrone fitting are consistent with that obtained from LAMOST spectra and binary modelling. According to the isochrone fitting, we find that the primary of KIC 8097825 somewhat leaves the main-sequence stage and the secondary is cose to the red giant branch. KIC 6859813 has two main-sequence stars. The primary of KIC 5527172 is on the main-sequence turn-off and the secondary is an M dwarf with inflation radius. 

The starspot signals are obtained by subtracting the binary model from the light curves. The RMSs, lifetimes, rotation periods, SDRs of starspots are determined by analyzing the starspot signals. Both components of KIC~8097825 have spots. Its rapidly rotating components present large starspots. The total duration time of single dips is longer than that of double. The primary of KIC~ 6859813 has spots. There are some clues indicating that the secondary also has spots. The rotation period of KIC~ 6859813 is shorter than its orbital period but its age is much longer than its synchronization timescale, which could be the result of the differential rotation at the surface. The total duration time of double dips is longer than that of single.  The primary of KIC~ 5527172 has spots. The total duration time of double dips is close to that of single. The spot lifetimes of KIC 8097825 and KIC 6859813 are consistent with those of single star systems, while KIC 5527172 has a long spot lifetime.

In order to characterize the sizes, lifetimes and SDRs of starspots between the binaries and single stars, the starspot modulation in DEBs with orbital solutions are collected from the literature and then compared with that in single stars. Our results show that the RMSs of starspots on binaries are lower than the medians of those on single stars, although most of them locate in the $1 \sigma$ region of those on single stars. The lifetimes of starspots are consistent with each other, when the rotation periods are larger than 3 days. Because of the lack of single star samples with shorter rotation periods, we have no idea of the situation in shorter period range. Binaries and single stars with shorter rotation periods are needed to investigate the tidal effect on the starspot evolution. SDRs for half of the binaries are consistent with those of single star systems, while another half are smaller. The relative lifetime positively correlates with the RMS and SDR but negatively correlates with the rotation period. Binaries with luminosity ratios close to the unit tend to have more double dips. The long-baseline high-quality multi-band photometric data is needed to obtain more details of starspots.

\section*{Acknowledgements}
JNF acknowledges the support from the National Natural Science Foundation of China (NSFC) through the grants 11833002, 12090040, and 12090042.
Guoshoujing Telescope (the Large Sky Area Multi-Object Fiber Spectroscopic Telescope LAMOST) is a National Major Scientific Project built by the Chinese Academy of Sciences. Funding for the project has been provided by the National Development and Reform Commission. LAMOST is operated and managed by the National Astronomical Observatories, Chinese Academy of Sciences. JXW hosts the LAMOST fellowship as a Youth Researcher which is supported by the Special Funding for Advanced Users, budgeted and administrated by the Center for Astronomical Mega-Science, Chinese Academy of Sciences (CAMS). JXW and JNF acknowledge the support from the Cultivation Project for LAMOST Scientific Payoff and Research Achievement of CAMS-CAS.
\addcontentsline{toc}{section}{Acknowledgements}


\section{DATA AVAILABILITY}
The data underlying this article will be shared on reasonable request
to the corresponding author.



\bibliographystyle{mnras}
\bibliography{Three_DEB} 

\begin{thebibliography}{}
\makeatletter
\relax
\def\mn@urlcharsother{\let\do\@makeother \do\$\do\&\do\#\do\^\do\_\do\%\do\~}
\def\mn@doi{\begingroup\mn@urlcharsother \@ifnextchar [ {\mn@doi@}
  {\mn@doi@[]}}
\def\mn@doi@[#1]#2{\def\@tempa{#1}\ifx\@tempa\@empty \href
  {http://dx.doi.org/#2} {doi:#2}\else \href {http://dx.doi.org/#2} {#1}\fi
  \endgroup}
\def\mn@eprint#1#2{\mn@eprint@#1:#2::\@nil}
\def\mn@eprint@arXiv#1{\href {http://arxiv.org/abs/#1} {{\tt arXiv:#1}}}
\def\mn@eprint@dblp#1{\href {http://dblp.uni-trier.de/rec/bibtex/#1.xml}
  {dblp:#1}}
\def\mn@eprint@#1:#2:#3:#4\@nil{\def\@tempa {#1}\def\@tempb {#2}\def\@tempc
  {#3}\ifx \@tempc \@empty \let \@tempc \@tempb \let \@tempb \@tempa \fi \ifx
  \@tempb \@empty \def\@tempb {arXiv}\fi \@ifundefined
  {mn@eprint@\@tempb}{\@tempb:\@tempc}{\expandafter \expandafter \csname
  mn@eprint@\@tempb\endcsname \expandafter{\@tempc}}}

\bibitem[\protect\citeauthoryear{{Aigrain} et~al.,}{{Aigrain}
  et~al.}{2015}]{2015MNRAS.450.3211A}
{Aigrain} S.,  et~al., 2015, \mn@doi [\mnras] {10.1093/mnras/stv853}, \href
  {https://ui.adsabs.harvard.edu/abs/2015MNRAS.450.3211A} {450, 3211}

\bibitem[\protect\citeauthoryear{{Barnes}, {Lister}, {Hilditch}  \& {Collier
  Cameron}}{{Barnes} et~al.}{2004}]{2004MNRAS.348.1321B}
{Barnes} J.~R.,  {Lister} T.~A.,  {Hilditch} R.~W.,   {Collier Cameron} A.,
  2004, \mn@doi [\mnras] {10.1111/j.1365-2966.2004.07452.x}, \href
  {https://ui.adsabs.harvard.edu/abs/2004MNRAS.348.1321B} {348, 1321}

\bibitem[\protect\citeauthoryear{{Basri} \& {Nguyen}}{{Basri} \&
  {Nguyen}}{2018}]{2018ApJ...863..190B}
{Basri} G.,  {Nguyen} H.~T.,  2018, \mn@doi [\apj] {10.3847/1538-4357/aad3b6},
  \href {https://ui.adsabs.harvard.edu/abs/2018ApJ...863..190B} {863, 190}

\bibitem[\protect\citeauthoryear{{Basri} \& {Shah}}{{Basri} \&
  {Shah}}{2020}]{2020ApJ...901...14B}
{Basri} G.,  {Shah} R.,  2020, \mn@doi [\apj] {10.3847/1538-4357/abae5d}, \href
  {https://ui.adsabs.harvard.edu/abs/2020ApJ...901...14B} {901, 14}

\bibitem[\protect\citeauthoryear{{Basri}, {Streichenberger}, {McWard},
  {Edmond}, {Tan}, {Lee}  \& {Melton}}{{Basri}
  et~al.}{2021}]{2021arXiv211013284B}
{Basri} G.,  {Streichenberger} T.,  {McWard} C.,  {Edmond} Lawrence I.,  {Tan}
  J.,  {Lee} M.,   {Melton} T.,  2021, arXiv e-prints, \href
  {https://ui.adsabs.harvard.edu/abs/2021arXiv211013284B} {p. arXiv:2110.13284}

\bibitem[\protect\citeauthoryear{{Berdyugina}}{{Berdyugina}}{2005}]{2005LRSP....2....8B}
{Berdyugina} S.~V.,  2005, \mn@doi [Living Reviews in Solar Physics]
  {10.12942/lrsp-2005-8}, \href
  {https://ui.adsabs.harvard.edu/abs/2005LRSP....2....8B} {2, 8}

\bibitem[\protect\citeauthoryear{{Borucki}}{{Borucki}}{2016}]{2016RPPh...79c6901B}
{Borucki} W.~J.,  2016, \mn@doi [Reports on Progress in Physics]
  {10.1088/0034-4885/79/3/036901}, \href
  {https://ui.adsabs.harvard.edu/abs/2016RPPh...79c6901B} {79, 036901}

\bibitem[\protect\citeauthoryear{{Bressan}, {Marigo}, {Girardi}, {Salasnich},
  {Dal Cero}, {Rubele}  \& {Nanni}}{{Bressan}
  et~al.}{2012}]{2012MNRAS.427..127B}
{Bressan} A.,  {Marigo} P.,  {Girardi} L.,  {Salasnich} B.,  {Dal Cero} C.,
  {Rubele} S.,   {Nanni} A.,  2012, \mn@doi [\mnras]
  {10.1111/j.1365-2966.2012.21948.x}, \href
  {https://ui.adsabs.harvard.edu/abs/2012MNRAS.427..127B} {427, 127}

\bibitem[\protect\citeauthoryear{{Clark Cunningham}, {Rawls}, {Windemuth},
  {Ali}, {Jackiewicz}, {Agol}  \& {Stassun}}{{Clark Cunningham}
  et~al.}{2019}]{2019AJ....158..106C}
{Clark Cunningham} J.~M.,  {Rawls} M.~L.,  {Windemuth} D.,  {Ali} A.,
  {Jackiewicz} J.,  {Agol} E.,   {Stassun} K.~G.,  2019, \mn@doi [\aj]
  {10.3847/1538-3881/ab2d2b}, \href
  {https://ui.adsabs.harvard.edu/abs/2019AJ....158..106C} {158, 106}

\bibitem[\protect\citeauthoryear{{Conroy} et~al.,}{{Conroy}
  et~al.}{2020}]{2020ApJS..250...34C}
{Conroy} K.~E.,  et~al., 2020, \mn@doi [\apjs] {10.3847/1538-4365/abb4e2},
  \href {https://ui.adsabs.harvard.edu/abs/2020ApJS..250...34C} {250, 34}

\bibitem[\protect\citeauthoryear{{Cui} et~al.,}{{Cui} et~al.}{2012}]{Cui2012}
{Cui} X.-Q.,  et~al., 2012, \mn@doi [Research in Astronomy and Astrophysics]
  {10.1088/1674-4527/12/9/003}, \href
  {https://ui.adsabs.harvard.edu/abs/2012RAA....12.1197C} {12, 1197}

\bibitem[\protect\citeauthoryear{{Czesla}, {Terzenbach}, {Wichmann}  \&
  {Schmitt}}{{Czesla} et~al.}{2019}]{2019A&A...623A.107C}
{Czesla} S.,  {Terzenbach} S.,  {Wichmann} R.,   {Schmitt} J.~H.~M.~M.,  2019,
  \mn@doi [\aap] {10.1051/0004-6361/201834516}, \href
  {https://ui.adsabs.harvard.edu/abs/2019A&A...623A.107C} {623, A107}

\bibitem[\protect\citeauthoryear{{De Cat} et~al.,}{{De Cat}
  et~al.}{2015}]{2015ApJS..220...19D}
{De Cat} P.,  et~al., 2015, \mn@doi [\apjs] {10.1088/0067-0049/220/1/19}, \href
  {https://ui.adsabs.harvard.edu/abs/2015ApJS..220...19D} {220, 19}

\bibitem[\protect\citeauthoryear{{El-Badry}, {Rix}, {Ting}, {Weisz},
  {Bergemann}, {Cargile}, {Conroy}  \& {Eilers}}{{El-Badry}
  et~al.}{2018}]{2018MNRAS.473.5043E}
{El-Badry} K.,  {Rix} H.-W.,  {Ting} Y.-S.,  {Weisz} D.~R.,  {Bergemann} M.,
  {Cargile} P.,  {Conroy} C.,   {Eilers} A.-C.,  2018, \mn@doi [\mnras]
  {10.1093/mnras/stx2758}, \href
  {https://ui.adsabs.harvard.edu/abs/2018MNRAS.473.5043E} {473, 5043}

\bibitem[\protect\citeauthoryear{{Fu} et~al.,}{{Fu}
  et~al.}{2020}]{2020RAA....20..167F}
{Fu} J.-N.,  et~al., 2020, \mn@doi [Research in Astronomy and Astrophysics]
  {10.1088/1674-4527/20/10/167}, \href
  {https://ui.adsabs.harvard.edu/abs/2020RAA....20..167F} {20, 167}

\bibitem[\protect\citeauthoryear{{Giles}, {Collier Cameron}  \&
  {Haywood}}{{Giles} et~al.}{2017}]{2017MNRAS.472.1618G}
{Giles} H. A.~C.,  {Collier Cameron} A.,   {Haywood} R.~D.,  2017, \mn@doi
  [\mnras] {10.1093/mnras/stx1931}, \href
  {https://ui.adsabs.harvard.edu/abs/2017MNRAS.472.1618G} {472, 1618}

\bibitem[\protect\citeauthoryear{{Gu}, {Tan}, {Wang}  \& {Shan}}{{Gu}
  et~al.}{2003}]{2003A&A...405..763G}
{Gu} S.~H.,  {Tan} H.~S.,  {Wang} X.~B.,   {Shan} H.~G.,  2003, \mn@doi [\aap]
  {10.1051/0004-6361:20030671}, \href
  {https://ui.adsabs.harvard.edu/abs/2003A&A...405..763G} {405, 763}

\bibitem[\protect\citeauthoryear{{Hambleton} et~al.,}{{Hambleton}
  et~al.}{2013}]{2013MNRAS.434..925H}
{Hambleton} K.~M.,  et~al., 2013, \mn@doi [\mnras] {10.1093/mnras/stt886},
  \href {https://ui.adsabs.harvard.edu/abs/2013MNRAS.434..925H} {434, 925}

\bibitem[\protect\citeauthoryear{{Han} et~al.,}{{Han}
  et~al.}{2017}]{2017AJ....154..100H}
{Han} E.,  et~al., 2017, \mn@doi [\aj] {10.3847/1538-3881/aa803c}, \href
  {https://ui.adsabs.harvard.edu/abs/2017AJ....154..100H} {154, 100}

\bibitem[\protect\citeauthoryear{{Han}, {Muirhead}  \& {Swift}}{{Han}
  et~al.}{2019}]{2019AJ....158..111H}
{Han} E.,  {Muirhead} P.~S.,   {Swift} J.~J.,  2019, \mn@doi [\aj]
  {10.3847/1538-3881/ab2ed7}, \href
  {https://ui.adsabs.harvard.edu/abs/2019AJ....158..111H} {158, 111}

\bibitem[\protect\citeauthoryear{{He{\l}miniak}, {Ukita}, {Kambe},
  {Koz{\l}owski}, {Sybilski}, {Ratajczak}, {Maehara}  \&
  {Konacki}}{{He{\l}miniak} et~al.}{2016}]{2016MNRAS.461.2896H}
{He{\l}miniak} K.~G.,  {Ukita} N.,  {Kambe} E.,  {Koz{\l}owski} S.~K.,
  {Sybilski} P.,  {Ratajczak} M.,  {Maehara} H.,   {Konacki} M.,  2016, \mn@doi
  [\mnras] {10.1093/mnras/stw1514}, \href
  {https://ui.adsabs.harvard.edu/abs/2016MNRAS.461.2896H} {461, 2896}

\bibitem[\protect\citeauthoryear{{He{\l}miniak} et~al.,}{{He{\l}miniak}
  et~al.}{2017}]{2017MNRAS.468.1726H}
{He{\l}miniak} K.~G.,  et~al., 2017, \mn@doi [\mnras] {10.1093/mnras/stx385},
  \href {https://ui.adsabs.harvard.edu/abs/2017MNRAS.468.1726H} {468, 1726}

\bibitem[\protect\citeauthoryear{{Hilditch}}{{Hilditch}}{2001}]{2001icbs.book.....H}
{Hilditch} R.~W.,  2001, {An Introduction to Close Binary Stars}

\bibitem[\protect\citeauthoryear{{Holzwarth} \& {Sch{\"u}ssler}}{{Holzwarth} \&
  {Sch{\"u}ssler}}{2000}]{2000AN....321..175H}
{Holzwarth} V.,  {Sch{\"u}ssler} M.,  2000, \mn@doi [Astronomische Nachrichten]
  {10.1002/1521-3994(200008)321:3\textless{}175::AID-ASNA175\textgreater{}3.0.CO;2-V},
  \href {https://ui.adsabs.harvard.edu/abs/2000AN....321..175H} {321, 175}

\bibitem[\protect\citeauthoryear{{Holzwarth} \& {Sch{\"u}ssler}}{{Holzwarth} \&
  {Sch{\"u}ssler}}{2002}]{2002AN....323..399H}
{Holzwarth} V.,  {Sch{\"u}ssler} M.,  2002, \mn@doi [Astronomische Nachrichten]
  {10.1002/1521-3994(200208)323:3/4\textless{}399::AID-ASNA399\textgreater{}3.0.CO;2-V},
  \href {https://ui.adsabs.harvard.edu/abs/2002AN....323..399H} {323, 399}

\bibitem[\protect\citeauthoryear{{Holzwarth} \& {Sch{\"u}ssler}}{{Holzwarth} \&
  {Sch{\"u}ssler}}{2003a}]{2003A&A...405..291H}
{Holzwarth} V.,  {Sch{\"u}ssler} M.,  2003a, \mn@doi [\aap]
  {10.1051/0004-6361:20030582}, \href
  {https://ui.adsabs.harvard.edu/abs/2003A&A...405..291H} {405, 291}

\bibitem[\protect\citeauthoryear{{Holzwarth} \& {Sch{\"u}ssler}}{{Holzwarth} \&
  {Sch{\"u}ssler}}{2003b}]{2003A&A...405..303H}
{Holzwarth} V.,  {Sch{\"u}ssler} M.,  2003b, \mn@doi [\aap]
  {10.1051/0004-6361:20030584}, \href
  {https://ui.adsabs.harvard.edu/abs/2003A&A...405..303H} {405, 303}

\bibitem[\protect\citeauthoryear{{Hussain}}{{Hussain}}{2002}]{2002AN....323..349H}
{Hussain} G.~A.~J.,  2002, \mn@doi [Astronomische Nachrichten]
  {10.1002/1521-3994(200208)323:3/4\textless{}349::AID-ASNA349\textgreater{}3.0.CO;2-E},
  \href {https://ui.adsabs.harvard.edu/abs/2002AN....323..349H} {323, 349}

\bibitem[\protect\citeauthoryear{{I{\c{s}}ik}, {Schmitt}  \&
  {Sch{\"u}ssler}}{{I{\c{s}}ik} et~al.}{2007a}]{2007AN....328.1111I}
{I{\c{s}}ik} E.,  {Schmitt} D.,   {Sch{\"u}ssler} M.,  2007a, \mn@doi
  [Astronomische Nachrichten] {10.1002/asna.200710865}, \href
  {https://ui.adsabs.harvard.edu/abs/2007AN....328.1111I} {328, 1111}

\bibitem[\protect\citeauthoryear{{I{\c{s}}ik}, {Sch{\"u}ssler}  \&
  {Solanki}}{{I{\c{s}}ik} et~al.}{2007b}]{2007A&A...464.1049I}
{I{\c{s}}ik} E.,  {Sch{\"u}ssler} M.,   {Solanki} S.~K.,  2007b, \mn@doi [\aap]
  {10.1051/0004-6361:20066623}, \href
  {https://ui.adsabs.harvard.edu/abs/2007A&A...464.1049I} {464, 1049}

\bibitem[\protect\citeauthoryear{{I{\c{s}}{\i}k}, {Schmitt}  \&
  {Sch{\"u}ssler}}{{I{\c{s}}{\i}k} et~al.}{2011}]{2011A&A...528A.135I}
{I{\c{s}}{\i}k} E.,  {Schmitt} D.,   {Sch{\"u}ssler} M.,  2011, \mn@doi [\aap]
  {10.1051/0004-6361/201014501}, \href
  {https://ui.adsabs.harvard.edu/abs/2011A&A...528A.135I} {528, A135}

\bibitem[\protect\citeauthoryear{{Jones} et~al.,}{{Jones}
  et~al.}{2020}]{2020ApJS..247...63J}
{Jones} D.,  et~al., 2020, \mn@doi [\apjs] {10.3847/1538-4365/ab7927}, \href
  {https://ui.adsabs.harvard.edu/abs/2020ApJS..247...63J} {247, 63}

\bibitem[\protect\citeauthoryear{{Kesseli}, {Muirhead}, {Mann}  \&
  {Mace}}{{Kesseli} et~al.}{2018}]{2018AJ....155..225K}
{Kesseli} A.~Y.,  {Muirhead} P.~S.,  {Mann} A.~W.,   {Mace} G.,  2018, \mn@doi
  [\aj] {10.3847/1538-3881/aabccb}, \href
  {https://ui.adsabs.harvard.edu/abs/2018AJ....155..225K} {155, 225}

\bibitem[\protect\citeauthoryear{{Kirk} et~al.,}{{Kirk}
  et~al.}{2016}]{2016AJ....151...68K}
{Kirk} B.,  et~al., 2016, \mn@doi [\aj] {10.3847/0004-6256/151/3/68}, \href
  {https://ui.adsabs.harvard.edu/abs/2016AJ....151...68K} {151, 68}

\bibitem[\protect\citeauthoryear{{Korhonen}, {Berdyugina}, {Hackman}, {Ilyin},
  {Strassmeier}  \& {Tuominen}}{{Korhonen} et~al.}{2007}]{2007A&A...476..881K}
{Korhonen} H.,  {Berdyugina} S.~V.,  {Hackman} T.,  {Ilyin} I.~V.,
  {Strassmeier} K.~G.,   {Tuominen} I.,  2007, \mn@doi [\aap]
  {10.1051/0004-6361:20041806}, \href
  {https://ui.adsabs.harvard.edu/abs/2007A&A...476..881K} {476, 881}

\bibitem[\protect\citeauthoryear{{K{\"u}ker} \& {R{\"u}diger}}{{K{\"u}ker} \&
  {R{\"u}diger}}{2011}]{2011AN....332..933K}
{K{\"u}ker} M.,  {R{\"u}diger} G.,  2011, \mn@doi [Astronomische Nachrichten]
  {10.1002/asna.201111628}, \href
  {https://ui.adsabs.harvard.edu/abs/2011AN....332..933K} {332, 933}

\bibitem[\protect\citeauthoryear{{Liu} et~al.,}{{Liu}
  et~al.}{2019}]{2019RAA....19...75L}
{Liu} N.,  et~al., 2019, \mn@doi [Research in Astronomy and Astrophysics]
  {10.1088/1674-4527/19/5/75}, \href
  {https://ui.adsabs.harvard.edu/abs/2019RAA....19...75L} {19, 075}

\bibitem[\protect\citeauthoryear{{Llama}, {Jardine}, {Mackay}  \&
  {Fares}}{{Llama} et~al.}{2012}]{2012MNRAS.422L..72L}
{Llama} J.,  {Jardine} M.,  {Mackay} D.~H.,   {Fares} R.,  2012, \mn@doi
  [\mnras] {10.1111/j.1745-3933.2012.01239.x}, \href
  {https://ui.adsabs.harvard.edu/abs/2012MNRAS.422L..72L} {422, L72}

\bibitem[\protect\citeauthoryear{{Luo}, {Zhang}  \& {Zhao}}{{Luo}
  et~al.}{2004}]{Luo2004}
{Luo} A.~L.,  {Zhang} Y.-X.,   {Zhao} Y.-H.,  2004, ] {10.1117/12.548737},
  \href {https://ui.adsabs.harvard.edu/abs/2004SPIE.5496..756L} {5496, 756}

\bibitem[\protect\citeauthoryear{{Luo} et~al.,}{{Luo}
  et~al.}{2012}]{2012RAA....12.1243L}
{Luo} A.~L.,  et~al., 2012, \mn@doi [Research in Astronomy and Astrophysics]
  {10.1088/1674-4527/12/9/004}, \href
  {https://ui.adsabs.harvard.edu/abs/2012RAA....12.1243L} {12, 1243}

\bibitem[\protect\citeauthoryear{{Lurie} et~al.,}{{Lurie}
  et~al.}{2017}]{2017AJ....154..250L}
{Lurie} J.~C.,  et~al., 2017, \mn@doi [\aj] {10.3847/1538-3881/aa974d}, \href
  {https://ui.adsabs.harvard.edu/abs/2017AJ....154..250L} {154, 250}

\bibitem[\protect\citeauthoryear{{Matson}, {Gies}, {Guo}  \&
  {Williams}}{{Matson} et~al.}{2017}]{2017AJ....154..216M}
{Matson} R.~A.,  {Gies} D.~R.,  {Guo} Z.,   {Williams} S.~J.,  2017, \mn@doi
  [\aj] {10.3847/1538-3881/aa8fd6}, \href
  {https://ui.adsabs.harvard.edu/abs/2017AJ....154..216M} {154, 216}

\bibitem[\protect\citeauthoryear{{McQuillan}, {Aigrain}  \&
  {Mazeh}}{{McQuillan} et~al.}{2013}]{2013MNRAS.432.1203M}
{McQuillan} A.,  {Aigrain} S.,   {Mazeh} T.,  2013, \mn@doi [\mnras]
  {10.1093/mnras/stt536}, \href
  {https://ui.adsabs.harvard.edu/abs/2013MNRAS.432.1203M} {432, 1203}

\bibitem[\protect\citeauthoryear{{McQuillan}, {Mazeh}  \&
  {Aigrain}}{{McQuillan} et~al.}{2014}]{2014ApJS..211...24M}
{McQuillan} A.,  {Mazeh} T.,   {Aigrain} S.,  2014, \mn@doi [\apjs]
  {10.1088/0067-0049/211/2/24}, \href
  {https://ui.adsabs.harvard.edu/abs/2014ApJS..211...24M} {211, 24}

\bibitem[\protect\citeauthoryear{{Namekata} et~al.,}{{Namekata}
  et~al.}{2020}]{2020ApJ...891..103N}
{Namekata} K.,  et~al., 2020, \mn@doi [\apj] {10.3847/1538-4357/ab7384}, \href
  {https://ui.adsabs.harvard.edu/abs/2020ApJ...891..103N} {891, 103}

\bibitem[\protect\citeauthoryear{{{\"O}zavc{\i}}, {{\c{S}}enavc{\i}},
  {I{\textcommabelow s}{\i}k}, {Hussain}, {O'Neal}, {Y{\i}lmaz}  \&
  {Selam}}{{{\"O}zavc{\i}} et~al.}{2018}]{2018MNRAS.474.5534O}
{{\"O}zavc{\i}} I.,  {{\c{S}}enavc{\i}} H.~V.,  {I{\textcommabelow s}{\i}k} E.,
   {Hussain} G.~A.~J.,  {O'Neal} D.,  {Y{\i}lmaz} M.,   {Selam} S.~O.,  2018,
  \mn@doi [\mnras] {10.1093/mnras/stx3053}, \href
  {https://ui.adsabs.harvard.edu/abs/2018MNRAS.474.5534O} {474, 5534}

\bibitem[\protect\citeauthoryear{{Pan}, {Fu}, {Zong}, {Zhang}, {Wang}  \&
  {Li}}{{Pan} et~al.}{2020}]{2020ApJ...905...67P}
{Pan} Y.,  {Fu} J.-N.,  {Zong} W.,  {Zhang} X.,  {Wang} J.,   {Li} C.,  2020,
  \mn@doi [\apj] {10.3847/1538-4357/abc250}, \href
  {https://ui.adsabs.harvard.edu/abs/2020ApJ...905...67P} {905, 67}

\bibitem[\protect\citeauthoryear{{Parsons} et~al.,}{{Parsons}
  et~al.}{2018}]{2018MNRAS.481.1083P}
{Parsons} S.~G.,  et~al., 2018, \mn@doi [\mnras] {10.1093/mnras/sty2345}, \href
  {https://ui.adsabs.harvard.edu/abs/2018MNRAS.481.1083P} {481, 1083}

\bibitem[\protect\citeauthoryear{Pedregosa et~al.,}{Pedregosa
  et~al.}{2011}]{JMLR:v12:pedregosa11a}
Pedregosa F.,  et~al., 2011, Journal of Machine Learning Research, 12, 2825

\bibitem[\protect\citeauthoryear{{Pi}, {Zhang}, {Bi}, {Han}, {Lu}, {Yue},
  {Long}  \& {Yan}}{{Pi} et~al.}{2019}]{2019ApJ...877...75P}
{Pi} Q.-f.,  {Zhang} L.-y.,  {Bi} S.-l.,  {Han} X.~L.,  {Lu} H.-p.,  {Yue} Q.,
  {Long} L.,   {Yan} Y.,  2019, \mn@doi [\apj] {10.3847/1538-4357/ab19c3},
  \href {https://ui.adsabs.harvard.edu/abs/2019ApJ...877...75P} {877, 75}

\bibitem[\protect\citeauthoryear{{Pr{\v{s}}a} \& {Zwitter}}{{Pr{\v{s}}a} \&
  {Zwitter}}{2005}]{2005ApJ...628..426P}
{Pr{\v{s}}a} A.,  {Zwitter} T.,  2005, \mn@doi [\apj] {10.1086/430591}, \href
  {https://ui.adsabs.harvard.edu/abs/2005ApJ...628..426P} {628, 426}

\bibitem[\protect\citeauthoryear{{Pr{\v{s}}a} et~al.,}{{Pr{\v{s}}a}
  et~al.}{2011}]{2011AJ....141...83P}
{Pr{\v{s}}a} A.,  et~al., 2011, \mn@doi [\aj] {10.1088/0004-6256/141/3/83},
  \href {https://ui.adsabs.harvard.edu/abs/2011AJ....141...83P} {141, 83}

\bibitem[\protect\citeauthoryear{{Pr{\v{s}}a} et~al.,}{{Pr{\v{s}}a}
  et~al.}{2016}]{2016ApJS..227...29P}
{Pr{\v{s}}a} A.,  et~al., 2016, \mn@doi [\apjs] {10.3847/1538-4365/227/2/29},
  \href {https://ui.adsabs.harvard.edu/abs/2016ApJS..227...29P} {227, 29}

\bibitem[\protect\citeauthoryear{{Rasmussen} \& {Williams}}{{Rasmussen} \&
  {Williams}}{2006}]{2006gpml.book.....R}
{Rasmussen} C.~E.,  {Williams} C. K.~I.,  2006, {Gaussian Processes for Machine
  Learning}

\bibitem[\protect\citeauthoryear{{Reinhold} \& {Gizon}}{{Reinhold} \&
  {Gizon}}{2015}]{2015A&A...583A..65R}
{Reinhold} T.,  {Gizon} L.,  2015, \mn@doi [\aap]
  {10.1051/0004-6361/201526216}, \href
  {https://ui.adsabs.harvard.edu/abs/2015A&A...583A..65R} {583, A65}

\bibitem[\protect\citeauthoryear{{Ren} et~al.,}{{Ren}
  et~al.}{2016}]{2016ApJS..225...28R}
{Ren} A.,  et~al., 2016, \mn@doi [\apjs] {10.3847/0067-0049/225/2/28}, \href
  {https://ui.adsabs.harvard.edu/abs/2016ApJS..225...28R} {225, 28}

\bibitem[\protect\citeauthoryear{{Santos}, {Mathur}, {Garc{\'\i}a}, {Cunha}  \&
  {Avelino}}{{Santos} et~al.}{2021}]{2021MNRAS.508..267S}
{Santos} A.~R.~G.,  {Mathur} S.,  {Garc{\'\i}a} R.~A.,  {Cunha} M.~S.,
  {Avelino} P.~P.,  2021, \mn@doi [\mnras] {10.1093/mnras/stab2402}, \href
  {https://ui.adsabs.harvard.edu/abs/2021MNRAS.508..267S} {508, 267}

\bibitem[\protect\citeauthoryear{{Slawson} et~al.,}{{Slawson}
  et~al.}{2011}]{2011AJ....142..160S}
{Slawson} R.~W.,  et~al., 2011, \mn@doi [\aj] {10.1088/0004-6256/142/5/160},
  \href {https://ui.adsabs.harvard.edu/abs/2011AJ....142..160S} {142, 160}

\bibitem[\protect\citeauthoryear{{Strassmeier}}{{Strassmeier}}{2009}]{2009A&ARv..17..251S}
{Strassmeier} K.~G.,  2009, \mn@doi [\aapr] {10.1007/s00159-009-0020-6}, \href
  {https://ui.adsabs.harvard.edu/abs/2009A&ARv..17..251S} {17, 251}

\bibitem[\protect\citeauthoryear{{Stumpe} et~al.,}{{Stumpe}
  et~al.}{2012}]{2012PASP..124..985S}
{Stumpe} M.~C.,  et~al., 2012, \mn@doi [\pasp] {10.1086/667698}, \href
  {https://ui.adsabs.harvard.edu/abs/2012PASP..124..985S} {124, 985}

\bibitem[\protect\citeauthoryear{{Su} \& {Cui}}{{Su} \&
  {Cui}}{2004}]{2004ChJAA...4....1S}
{Su} D.-Q.,  {Cui} X.-Q.,  2004, \mn@doi [\cjaa] {10.1088/1009-9271/4/1/1},
  \href {https://ui.adsabs.harvard.edu/abs/2004ChJAA...4....1S} {4, 1}

\bibitem[\protect\citeauthoryear{{Wang}, {Su}, {Chu}, {Cui}  \& {Wang}}{{Wang}
  et~al.}{1996}]{1996ApOpt..35.5155W}
{Wang} S.-G.,  {Su} D.-Q.,  {Chu} Y.-Q.,  {Cui} X.,   {Wang} Y.-N.,  1996,
  \mn@doi [\ao] {10.1364/AO.35.005155}, \href
  {https://ui.adsabs.harvard.edu/abs/1996ApOpt..35.5155W} {35, 5155}

\bibitem[\protect\citeauthoryear{{Wang}, {Fu}, {Niu}, {Pan}, {Li}, {Zong}  \&
  {Hou}}{{Wang} et~al.}{2021}]{2021MNRAS.504.4302W}
{Wang} J.,  {Fu} J.,  {Niu} H.,  {Pan} Y.,  {Li} C.,  {Zong} W.,   {Hou} Y.,
  2021, \mn@doi [\mnras] {10.1093/mnras/stab1219}, \href
  {https://ui.adsabs.harvard.edu/abs/2021MNRAS.504.4302W} {504, 4302}

\bibitem[\protect\citeauthoryear{{Wilson} \& {Devinney}}{{Wilson} \&
  {Devinney}}{1971}]{1971ApJ...166..605W}
{Wilson} R.~E.,  {Devinney} E.~J.,  1971, \mn@doi [\apj] {10.1086/150986},
  \href {https://ui.adsabs.harvard.edu/abs/1971ApJ...166..605W} {166, 605}

\bibitem[\protect\citeauthoryear{Wu, Singh, Prugniel, Gupta  \& Koleva}{Wu
  et~al.}{2010}]{Wu2010}
Wu Y.,  Singh H.~P.,  Prugniel P.,  Gupta R.,   Koleva M.,  2010, \mn@doi
  [Astronomy and Astrophysics] {10.1051/0004-6361/201015014}, 525, 1

\bibitem[\protect\citeauthoryear{Wu, Du, Luo, Zhao  \& Yuan}{Wu
  et~al.}{2014}]{Wu2014}
Wu Y.,  Du B.,  Luo A.,  Zhao Y.,   Yuan H.,  2014, \mn@doi [Proceedings of the
  International Astronomical Union] {10.1017/S1743921314010825}, 10, 340

\bibitem[\protect\citeauthoryear{{Xiang} et~al.,}{{Xiang}
  et~al.}{2020a}]{2020MNRAS.492.3647X}
{Xiang} Y.,  et~al., 2020a, \mn@doi [\mnras] {10.1093/mnras/staa063}, \href
  {https://ui.adsabs.harvard.edu/abs/2020MNRAS.492.3647X} {492, 3647}

\bibitem[\protect\citeauthoryear{{Xiang} et~al.,}{{Xiang}
  et~al.}{2020b}]{2020ApJ...893..164X}
{Xiang} Y.,  et~al., 2020b, \mn@doi [\apj] {10.3847/1538-4357/ab8229}, \href
  {https://ui.adsabs.harvard.edu/abs/2020ApJ...893..164X} {893, 164}

\bibitem[\protect\citeauthoryear{{Xu}, {Gu}  \& {Ioannidis}}{{Xu}
  et~al.}{2021}]{2021MNRAS.501.1878X}
{Xu} F.,  {Gu} S.,   {Ioannidis} P.,  2021, \mn@doi [\mnras]
  {10.1093/mnras/staa3793}, \href
  {https://ui.adsabs.harvard.edu/abs/2021MNRAS.501.1878X} {501, 1878}

\bibitem[\protect\citeauthoryear{{Yadav}, {Gastine}, {Christensen}  \&
  {Reiners}}{{Yadav} et~al.}{2015}]{2015A&A...573A..68Y}
{Yadav} R.~K.,  {Gastine} T.,  {Christensen} U.~R.,   {Reiners} A.,  2015,
  \mn@doi [\aap] {10.1051/0004-6361/201424589}, \href
  {https://ui.adsabs.harvard.edu/abs/2015A&A...573A..68Y} {573, A68}

\bibitem[\protect\citeauthoryear{{Zhang} et~al.,}{{Zhang}
  et~al.}{2021}]{2021ApJS..256...14Z}
{Zhang} B.,  et~al., 2021, \mn@doi [\apjs] {10.3847/1538-4365/ac0834}, \href
  {https://ui.adsabs.harvard.edu/abs/2021ApJS..256...14Z} {256, 14}

\bibitem[\protect\citeauthoryear{{Zhao}, {Zhao}, {Chu}, {Jing}  \&
  {Deng}}{{Zhao} et~al.}{2012}]{2012RAA....12..723Z}
{Zhao} G.,  {Zhao} Y.-H.,  {Chu} Y.-Q.,  {Jing} Y.-P.,   {Deng} L.-C.,  2012,
  \mn@doi [Research in Astronomy and Astrophysics]
  {10.1088/1674-4527/12/7/002}, \href
  {https://ui.adsabs.harvard.edu/abs/2012RAA....12..723Z} {12, 723}

\bibitem[\protect\citeauthoryear{{Zong} et~al.,}{{Zong}
  et~al.}{2018}]{2018ApJS..238...30Z}
{Zong} W.,  et~al., 2018, \mn@doi [\apjs] {10.3847/1538-4365/aadf81}, \href
  {https://ui.adsabs.harvard.edu/abs/2018ApJS..238...30Z} {238, 30}

\bibitem[\protect\citeauthoryear{{Zong} et~al.,}{{Zong}
  et~al.}{2020}]{2020ApJS..251...15Z}
{Zong} W.,  et~al., 2020, \mn@doi [\apjs] {10.3847/1538-4365/abbb2d}, \href
  {https://ui.adsabs.harvard.edu/abs/2020ApJS..251...15Z} {251, 15}

\bibitem[\protect\citeauthoryear{{{\c{S}}enavc{\i}}, {Hussain}, {O'Neal}  \&
  {Barnes}}{{{\c{S}}enavc{\i}} et~al.}{2011}]{2011A&A...529A..11S}
{{\c{S}}enavc{\i}} H.~V.,  {Hussain} G.~A.~J.,  {O'Neal} D.,   {Barnes} J.~R.,
  2011, \mn@doi [\aap] {10.1051/0004-6361/201015753}, \href
  {https://ui.adsabs.harvard.edu/abs/2011A&A...529A..11S} {529, A11}

\makeatother
\end{thebibliography}



\appendix

\renewcommand\thetable{\Alph{section}\arabic{table}}    

\section{Radial velocity measurements}
\setcounter{table}{0}

\begin{center}
	\begin{table*}
		\center \caption{ Radial velocity of KIC 8097825}
		\label{tab:rv1}
		\begin{tabular}{llccccc} 
			\hline
			&      &  \multicolumn{2}{|c|}{Primary}         &    & \multicolumn{2}{|c|}{Secondary}    \\
			\cline{3-4} \cline{6-7} \\
			BJD-2450000 & phase         & RV         & Residual       &   & RV  & Residual  \\
		
			&          & ${\rm (km/s)}$   & ${\rm (km/s)}$       &   & ${\rm (km/s)}$ & ${\rm (km/s)}$  \\
			
			\hline

			8267.231159 & 0.85319 & $24.1 \pm 3.4$ & -3.8 & & $-125.9 \pm 4.8$ & -0.3\\
			8267.243659 & 0.85744 & $21.7 \pm 4.6$ & -4.4 & & $-125.2 \pm 3.0$ & -1.1\\
			8267.256854 & 0.86193 & $19.9 \pm 4.7$ & -4.4 & & $-123.7 \pm 4.4$ & -1.2\\
			8267.269354 & 0.86619 & $20.4 \pm 3.7$ & -2.1 & & $-120.6 \pm 2.2$ & 0.3\\
			8267.281855 & 0.87045 & $18.8 \pm 2.4$ & -1.8 & & $-118.8 \pm 3.5$ & 0.5\\
			8267.295050 & 0.87494 & $18.2 \pm 3.7$ & -0.5 & & $-116.9 \pm 4.8$ & 0.6\\
			8267.315189 & 0.88180 & $13.4 \pm 9.4$ & -2.1 & & $-113.8 \pm 2.6$ & 0.9\\
			8268.238835 & 0.19630 & $-151.3 \pm 9.1$ & -0.9 & & $33.5 \pm 1.8$ & 2.3\\
			8268.248557 & 0.19961 & $-153.8 \pm 4.6$ & -2.8 & & $29.2 \pm 1.5$ & -2.6\\
			8268.257585 & 0.20268 & $-156.4 \pm 7.3$ & -4.8 & & $32.9 \pm 8.5$ & 0.5\\
			8268.266614 & 0.20576 & $-155.0 \pm 2.5$ & -2.8 & & $31.3 \pm 2.7$ & -1.6\\
			8268.276336 & 0.20907 & $-156.7 \pm 3.4$ & -3.9 & & $31.9 \pm 3.9$ & -1.5\\
			8268.286059 & 0.21238 & $-159.9 \pm 2.8$ & -6.6 & & $32.7 \pm 3.3$ & -1.2\\
			8268.295087 & 0.21545 & $-153.7 \pm 3.8$ & 0.0 & & $34.5 \pm 1.6$ & 0.3\\
			8268.316615 & 0.22278 & $-159.5 \pm 5.7$ & -4.9 & & $36.5 \pm 2.2$ & 1.5\\
			8269.269428 & 0.54722 & $-28.8 \pm 2.9$ & -4.9 & & $-71.8 \pm 1.7$ & 8.3\\
			8269.295124 & 0.55597 & $-17.2 \pm 3.5$ & 1.3 & & $-80.4 \pm 4.9$ & 4.5\\
			8269.308318 & 0.56046 & $-23.7 \pm 4.7$ & -7.9 & & $-84.2 \pm 2.8$ & 2.9\\
			8270.260436 & 0.88466 & $11.6 \pm 3.1$ & -2.5 & & $-111.6 \pm 3.8$ & 1.9\\
			8270.276409 & 0.89010 & $10.8 \pm 6.1$ & -0.6 & & $-111.1 \pm 2.1$ & 0.0\\
			8270.309049 & 0.90121 & $5.5 \pm 3.8$ & -0.3 & & $-105.8 \pm 2.6$ & 0.4\\
			8625.278099 & 0.76839 & $52.8 \pm 6.8$ & 4.9 & & $-141.2 \pm 3.8$ & 2.1\\
			8625.294072 & 0.77383 & $51.1 \pm 3.4$ & 3.7 & & $-141.7 \pm 5.6$ & 1.1\\
			8625.310739 & 0.77950 & $32.2 \pm 2.1$ & -14.6 & & $-144.7 \pm 1.2$ & -2.4\\
			8644.232259 & 0.22229 & $-156.5 \pm 4.9$ & -1.9 & & $38.4 \pm 3.5$ & 3.4\\
			8644.248926 & 0.22797 & $-156.1 \pm 2.7$ & -1.0 & & $37.6 \pm 4.0$ & 2.1\\
			8644.264899 & 0.23340 & $-156.5 \pm 4.2$ & -0.9 & & $39.4 \pm 2.8$ & 3.5\\
			8644.281566 & 0.23908 & $-156.5 \pm 3.5$ & -0.6 & & $38.5 \pm 2.2$ & 2.4\\
			8646.251768 & 0.90993 & $1.3 \pm 2.7$ & 0.2 & & $-96.4 \pm 1.8$ & 5.6\\
			8646.268435 & 0.91561 & $-1.5 \pm 3.0$ & 0.5 & & $-93.7 \pm 2.2$ & 5.6\\
			8646.290658 & 0.92318 & $-6.6 \pm 3.0$ & -0.3 & & $-90.9 \pm 9.2$ & 4.7\\
			8649.238666 & 0.92697 & $-11.1 \pm 3.4$ & -2.7 & & $-85.9 \pm 4.6$ & 7.8\\
			9001.232667 & 0.78115 & $47.6 \pm 5.7$ & 0.9 & & $-143.0 \pm 2.7$ & -0.9\\
			9001.278502 & 0.79675 & $47.2 \pm 4.6$ & 3.0 & & $-140.0 \pm 2.9$ & 0.0\\
			9001.294475 & 0.80219 & $46.2 \pm 4.3$ & 3.0 & & $-140.0 \pm 2.8$ & -0.9\\
			9001.311142 & 0.80787 & $44.5 \pm 6.1$ & 2.6 & & $-137.4 \pm 1.8$ & 0.5\\
			9004.244579 & 0.80670 & $47.0 \pm 3.9$ & 4.9 & & $-137.7 \pm 2.3$ & 0.5\\
			9004.276524 & 0.81758 & $41.4 \pm 3.6$ & 1.9 & & $-133.7 \pm 3.6$ & 2.2\\
			9004.293192 & 0.82326 & $41.6 \pm 4.7$ & 3.6 & & $-133.7 \pm 4.2$ & 0.7\\
			9004.309164 & 0.82870 & $41.5 \pm 4.2$ & 5.1 & & $-132.1 \pm 3.2$ & 1.0\\
			9011.262174 & 0.19620 & $-155.9 \pm 3.6$ & -5.6 & & $31.7 \pm 2.9$ & 0.5\\
			9011.278841 & 0.20188 & $-155.0 \pm 4.5$ & -3.5 & & $36.6 \pm 5.3$ & 4.4\\
			9011.311481 & 0.21299 & $-158.2 \pm 2.5$ & -4.9 & & $38.5 \pm 2.0$ & 4.6\\
			9015.261602 & 0.55801 & $-24.1 \pm 9.4$ & -6.9 & & $-83.1 \pm 2.5$ & 2.8\\
			9015.278269 & 0.56368 & $-18.0 \pm 3.7$ & -4.1 & & $-84.5 \pm 8.4$ & 4.4\\
			9015.294242 & 0.56912 & $-16.5 \pm 2.9$ & -5.8 & & $-89.6 \pm 2.7$ & 2.1\\
			9016.249826 & 0.89450 & $9.4 \pm 3.9$ & 0.2 & & $-108.2 \pm 3.4$ & 1.1\\
			9016.265799 & 0.89994 & $4.3 \pm 3.4$ & -2.1 & & $-105.6 \pm 5.3$ & 1.1\\
			9016.298438 & 0.91105 & $-1.7 \pm 3.7$ & -2.2 & & $-101.7 \pm 5.6$ & -0.2\\

			\hline
		\end{tabular}
	\end{table*}
\end{center}

\begin{center}
	\begin{table*}

	\center \caption{ Radial velocity of KIC 6859813}
	\label{tab:rv2}
	
	\begin{tabular}{llccccc} 
		\hline
		&      &  \multicolumn{2}{|c|}{Primary}         &    & \multicolumn{2}{|c|}{Secondary}    \\
		\cline{3-4} \cline{6-7}
		BJD-2450000 & phase         & RV         & Residual       &   & RV  & Residual  \\
		
		&          & ${\rm (km/s)}$   & ${\rm (km/s)}$       &   & ${\rm (km/s)}$ & ${\rm (km/s)}$  \\
		
		\hline
	
		8625.287291 & 0.28646 & $-74.5 \pm 0.6$ & 3.1 & & $-14.3 \pm 1.3$ & -2.1\\
		8644.225489 & 0.02671 & $-89.8 \pm 0.7$ & -3.9 & & $-1.5 \pm 2.2$ & 1.8\\
		8644.242156 & 0.02825 & $-83.6 \pm 0.6$ & 3.7 & & $-0.4 \pm 1.1$ & 1.3\\
		8644.258129 & 0.02971 & $-85.9 \pm 0.7$ & 2.7 & & $-0.0 \pm 1.1$ & 0.3\\
		8644.274101 & 0.03118 & $-89.2 \pm 0.8$ & 0.6 & & $0.7 \pm 0.9$ & -0.3\\
		8646.260971 & 0.21376 & $-99.8 \pm 0.7$ & 0.3 & & $10.1 \pm 1.7$ & -2.1\\
		8646.283888 & 0.21586 & $-102.0 \pm 0.7$ & -2.6 & & $16.8 \pm 1.8$ & 5.4\\
		9001.225895 & 0.83195 & $3.1 \pm 0.7$ & -0.1 & & $-101.7 \pm 0.7$ & -1.5\\
		9001.253673 & 0.83450 & $3.0 \pm 0.7$ & -0.1 & & $-100.1 \pm 0.4$ & -0.1\\
		9001.271730 & 0.83616 & $2.8 \pm 0.7$ & -0.2 & & $-99.4 \pm 0.4$ & 0.6\\
		9001.287702 & 0.83763 & $2.5 \pm 0.7$ & -0.5 & & $-101.7 \pm 0.6$ & -1.8\\
		9001.304370 & 0.83916 & $4.0 \pm 0.7$ & 1.2 & & $-97.5 \pm 0.8$ & 2.3\\
		9003.219716 & 0.01516 & $-73.8 \pm 0.6$ & 1.6 & & $-14.0 \pm 0.4$ & 0.7\\
		9003.235689 & 0.01663 & $-75.2 \pm 0.6$ & 1.5 & & $-12.2 \pm 0.5$ & 1.0\\
		9003.251662 & 0.01810 & $-77.7 \pm 0.6$ & 0.4 & & $-10.3 \pm 0.6$ & 1.4\\
		9003.268329 & 0.01963 & $-78.0 \pm 0.6$ & 1.4 & & $-9.5 \pm 0.4$ & 0.7\\
		9003.284302 & 0.02110 & $-79.6 \pm 0.7$ & 1.2 & & $-8.7 \pm 0.4$ & 0.1\\
		9003.300275 & 0.02256 & $-81.3 \pm 0.7$ & 0.9 & & $-7.1 \pm 0.4$ & 0.2\\
		9004.237808 & 0.10872 & $-126.1 \pm 0.7$ & -0.5 & & $39.4 \pm 0.4$ & -0.4\\
		9004.253781 & 0.11018 & $-126.0 \pm 0.7$ & -0.5 & & $39.9 \pm 0.4$ & 0.1\\
		9004.269754 & 0.11165 & $-125.9 \pm 0.7$ & -0.4 & & $40.8 \pm 0.4$ & 1.0\\
		9004.286421 & 0.11318 & $-125.6 \pm 0.7$ & -0.1 & & $39.2 \pm 0.4$ & -0.6\\
		9004.302394 & 0.11465 & $-125.5 \pm 0.7$ & -0.1 & & $39.9 \pm 0.4$ & 0.2\\
		9006.291351 & 0.29742 & $-74.6 \pm 0.7$ & -0.1 & & $-15.1 \pm 0.4$ & 0.5\\
		9011.255404 & 0.75357 & $1.2 \pm 0.7$ & -0.7 & & $-99.5 \pm 0.5$ & -0.7\\
		9011.272071 & 0.75510 & $0.3 \pm 0.6$ & -1.7 & & $-101.4 \pm 0.6$ & -2.5\\
		9011.288044 & 0.75657 & $1.3 \pm 0.7$ & -0.7 & & $-98.3 \pm 0.9$ & 0.7\\
		9011.304711 & 0.75810 & $3.0 \pm 0.7$ & 0.8 & & $-99.7 \pm 0.9$ & -0.7\\
		9015.199275 & 0.11598 & $-125.7 \pm 0.7$ & -0.4 & & $40.3 \pm 0.4$ & 0.7\\
		9015.215247 & 0.11745 & $-126.2 \pm 0.7$ & -1.0 & & $38.7 \pm 0.4$ & -0.8\\
		9015.231915 & 0.11898 & $-125.7 \pm 0.7$ & -0.6 & & $39.1 \pm 0.4$ & -0.3\\
		9015.254832 & 0.12108 & $-125.7 \pm 0.7$ & -0.8 & & $39.0 \pm 0.4$ & -0.1\\
		9015.271499 & 0.12262 & $-125.2 \pm 0.7$ & -0.4 & & $38.4 \pm 0.4$ & -0.5\\
		9015.287472 & 0.12408 & $-124.8 \pm 0.7$ & -0.2 & & $38.3 \pm 0.5$ & -0.5\\
		9016.242361 & 0.21183 & $-102.1 \pm 0.7$ & -1.3 & & $12.6 \pm 0.5$ & -0.3\\
		9016.259028 & 0.21336 & $-101.0 \pm 0.7$ & -0.7 & & $11.6 \pm 0.5$ & -0.7\\
		9016.275001 & 0.21483 & $-100.2 \pm 0.6$ & -0.4 & & $10.8 \pm 0.4$ & -1.0\\
		9016.291668 & 0.21636 & $-99.9 \pm 0.7$ & -0.6 & & $11.7 \pm 0.6$ & 0.5\\
		\hline
	\end{tabular}
\end{table*}
\end{center}

\begin{center}
	\begin{table*}
		\center \caption{ Radial velocity of KIC 5527172}
		\label{tab:rv1}
		\begin{tabular}{llcc} 
			\hline

			BJD-2450000 & phase         & RV               & Residual         \\
			
			    &               & ${\rm (km/s)}$   & ${\rm (km/s)}$      \\
			
			\hline
			
			8267.226634 & 0.87139 & $-26.6 \pm 0.9$ & 0.0\\
			8267.239134 & 0.87532 & $-28.2 \pm 0.9$ & -0.9\\
			8267.252329 & 0.87946 & $-28.2 \pm 0.9$ & -0.2\\
			8267.264830 & 0.88339 & $-28.2 \pm 1.0$ & 0.4\\
			8267.277330 & 0.88731 & $-28.8 \pm 0.9$ & 0.5\\
			8267.290525 & 0.89146 & $-30.4 \pm 1.0$ & -0.4\\
			8267.310665 & 0.89778 & $-30.4 \pm 1.0$ & 0.7\\
			8268.235700 & 0.18831 & $-82.2 \pm 0.9$ & -0.6\\
			8268.245423 & 0.19137 & $-83.1 \pm 0.9$ & -1.2\\
			8268.254451 & 0.19420 & $-83.0 \pm 0.9$ & -1.0\\
			8268.263479 & 0.19704 & $-83.0 \pm 0.9$ & -0.7\\
			8268.273202 & 0.20009 & $-83.7 \pm 0.9$ & -1.2\\
			8268.282924 & 0.20315 & $-85.0 \pm 0.9$ & -2.2\\
			8268.291952 & 0.20598 & $-83.3 \pm 1.0$ & -0.4\\
			8268.303758 & 0.20969 & $-84.2 \pm 0.9$ & -1.1\\
			8268.313481 & 0.21274 & $-84.0 \pm 0.9$ & -0.7\\
			8269.251016 & 0.50720 & $-48.1 \pm 0.8$ & -0.1\\
			8269.264906 & 0.51156 & $-47.2 \pm 0.8$ & -0.3\\
			8269.278101 & 0.51570 & $-45.9 \pm 0.8$ & -0.0\\
			8269.290601 & 0.51963 & $-45.3 \pm 0.8$ & -0.3\\
			8269.303796 & 0.52377 & $-44.3 \pm 0.8$ & -0.3\\
			8270.253832 & 0.82216 & $-19.0 \pm 0.8$ & 0.8\\
			8270.269804 & 0.82717 & $-19.9 \pm 0.8$ & 0.5\\
			8270.286472 & 0.83241 & $-20.4 \pm 0.8$ & 0.6\\
			8270.302445 & 0.83742 & $-21.5 \pm 0.8$ & 0.1\\
			8644.225659 & 0.27685 & $-82.9 \pm 0.8$ & 1.5\\
			8644.242326 & 0.28209 & $-84.1 \pm 0.8$ & 0.1\\
			8644.258299 & 0.28710 & $-84.0 \pm 0.8$ & 0.0\\
			8644.274271 & 0.29212 & $-83.1 \pm 0.8$ & 0.7\\
			8646.245168 & 0.91113 & $-37.1 \pm 1.4$ & -3.6\\
			8646.261141 & 0.91614 & $-35.6 \pm 0.9$ & -1.1\\
			8646.284058 & 0.92334 & $-38.2 \pm 1.0$ & -2.4\\
			9001.226063 & 0.40127 & $-67.7 \pm 1.4$ & 3.4\\
			9016.242530 & 0.11755 & $-71.3 \pm 0.9$ & 1.8\\
			9016.259197 & 0.12278 & $-72.3 \pm 0.8$ & 1.5\\
			9016.275170 & 0.12780 & $-72.6 \pm 0.8$ & 2.0\\
			9016.291837 & 0.13303 & $-74.1 \pm 0.8$ & 1.2\\
			\hline
		\end{tabular}
	\end{table*}
\end{center}

\section{The segments of starspot signals}

\begin{figure*}
	\centering
	
	\includegraphics[scale=0.5]{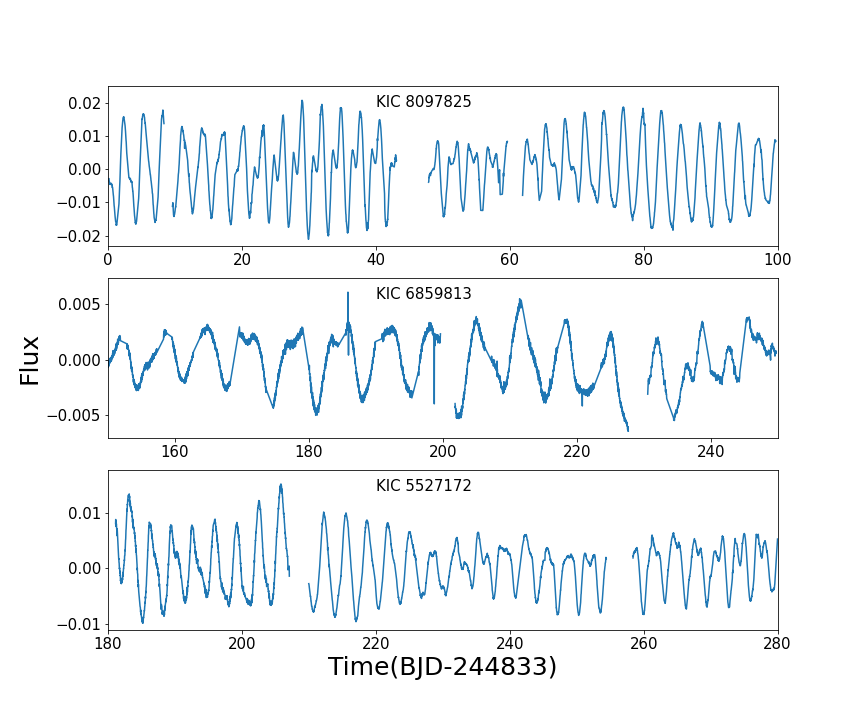}
	\caption{The segments of starspot signals of KIC 8097825, KIC 6859813, and KIC 5527172. These segments are chosen to show the single- and double-dip patterns.}
	\label{fig:spot_segment} 
\end{figure*}

\bsp	
\label{lastpage}
\end{document}